\documentclass[aps,floats]{revtex4}
\usepackage{amsmath,amssymb}
\usepackage{graphicx,epsfig}
\usepackage[greek,english]{babel}
\usepackage{bbold}

\usepackage{hyperref}

\begin{document}
\bibliographystyle {plain}
\pdfoutput=1
\def\oppropto{\mathop{\propto}} 
\def\opsimeq{\mathop{\simeq}}
\def\opoverderline{\mathop{\overline}}
\def\operarrow{\mathop{\longrightarrow}}
\def\opsim{\mathop{\sim}}

\def\opmin{\mathop{\min}} 
\def\opmax{\mathop{\max}} 
\def\oplim{\mathop{\lim}}

\def\fig#1#2{\includegraphics[height=#1]{#2}}
\def\figx#1#2{\includegraphics[width=#1]{#2}}


\title{ Jump-Drift and Jump-Diffusion Processes : Large Deviations \\
for the density, the current and the jump-flow and for the excursions between jumps} 


\author{ C\'ecile Monthus }
 \affiliation{Institut de Physique Th\'{e}orique, 
Universit\'e Paris Saclay, CNRS, CEA,
91191 Gif-sur-Yvette, France}

\begin{abstract}
For one-dimensional Jump-Drift and Jump-Diffusion processes converging towards some steady state, the large deviations of a long dynamical trajectory are described from two perspectives. Firstly, the joint probability of the empirical time-averaged density, of the empirical time-averaged current and of the empirical time-averaged jump-flow are studied via the large deviations at Level 2.5. Secondly, the joint probability of the empirical jumps and of the empirical excursions between consecutive jumps are analyzed via the large deviations at Level 2.5 for the alternate Markov chain that governs the series of all the jump events of a long trajectory. These two general frameworks are then applied to three examples of positive jump-drift processes without diffusion, and to two examples of jump-diffusion processes, in order to illustrate various simplifications that may occur in rate functions and in contraction procedures.

\end{abstract}

\maketitle


\section{ Introduction }

Jump-drift and jump-diffusion processes play a major role in many applications, in particular 
in mathematical finance \cite{book_finance,review_finance}), in biology for integrate-and-fire neuronal models \cite{dumont,miles} and in ecology to describe fires \cite{daly_fire} or soil moisture \cite{waterbalance,daly_rain,daly_rainbis,rainfall}.
They have also attracted a lot of interest in the field of intermittent search strategies 
(see the review \cite{review_search} and references therein)
and in the context of stochastic resetting (see the review \cite{review_reset} and references therein).

 In the present paper, the goal is to analyze the large deviations (see the reviews \cite{oono,ellis,review_touchette} and references therein)
 of the one-dimensional jump-drift or jump-diffusion process, whenever the space-dependent parameters of the model, namely the drift $v(x)$, the diffusion coefficient $D(x)$, the jump rate $\lambda(x)$ and the jump probability $\Pi(x' \vert x)$ (see section \ref{sec_diff})
 produce a localized steady state.
 
On one hand, the fluctuations of the empirical time-averaged density, 
of the empirical time-averaged current and of the empirical time-averaged jump-flow of a long dynamical trajectory
can be analyzed using the explicit large deviations at Level 2.5 for both 
continuous-time Markov Jump processes
\cite{fortelle_thesis,fortelle_jump,maes_canonical,maes_onandbeyond,wynants_thesis,chetrite_formal,BFG1,BFG2,chetrite_HDR,c_ring,c_interactions,c_open,c_detailed,barato_periodic,chetrite_periodic,c_reset,c_inference}
and for Diffusion processes 
\cite{wynants_thesis,maes_diffusion,chetrite_formal,engel,chetrite_HDR,c_reset,c_lyapunov,c_inference}.
This Level 2.5 can be then contracted to obtain the large deviations properties of any time-additive observable
of the dynamical trajectory. The link with the studies of time-additive observables 
via deformed Markov operators  \cite{derrida-lecture,sollich_review,lazarescu_companion,lazarescu_generic,jack_review,vivien_thesis,lecomte_chaotic,lecomte_thermo,lecomte_formalism,lecomte_glass,kristina1,kristina2,jack_ensemble,simon1,simon2,simon3,Gunter1,Gunter2,Gunter3,Gunter4,chetrite_canonical,chetrite_conditioned,chetrite_optimal,chetrite_HDR,touchette_circle,touchette_langevin,touchette_occ,touchette_occupation,derrida-conditioned,derrida-ring,bertin-conditioned,touchette-reflected,touchette-reflectedbis,c_lyapunov,previousquantum2.5doob,quantum2.5doob,quantum2.5dooblong,c_ruelle}
 can be understood via the corresponding 'conditioned' process obtained from the generalization of Doob's h-transform.

On the other hand, the series of all the jump events of a long trajectory can be analyzed via 
the alternate Markov chain that governs the jumps and the excursions between consecutive jumps.
As a consequence, the fluctuations of the empirical jump events and of the empirical excursions between consecutive jumps can be derived from the large deviations at Level 2.5 for discrete-time Markov chains 
 \cite{fortelle_thesis,fortelle_chain,review_touchette,c_largedevdisorder,c_reset,c_inference}.
 The large deviations for excusions between jumps have been studied previously 
for the special case of stochastic resetting to the origin \cite{touchette2015,maj2019,c_reset} 
and for run-and-tumble processes \cite{c_runandtumble}.

The paper is organized as follows.
In section \ref{sec_diff}, the general one-dimensional jump-diffusion model is defined in terms of four space-dependent parameters, namely the drift $v(x)$, the diffusion coefficient $D(x)$, the jump rate $\lambda(x)$ and the jump probability $\Pi(x' \vert x)$.
In section \ref{sec_2.5}, the large deviations at level 2.5 are analyzed for the 
joint distribution of the empirical density $\rho(x) $, of the empirical current $j(x)$ and of the empirical jump-flow $Q(x,y) $.
In section \ref{sec_contractionQflow}, the contraction over the jump-flow $Q(x,y) $ for given in/out-flows $Q^{\pm}(.)$ is analyzed and explicit solutions are given for two simple cases.
In section \ref{sec_jumpsonly}, the jump-diffusion dynamics is analyzed from the point of view of the jump events and of the excursions between the consecutive jumps.
In section \ref{sec_largedevexcursion}, the large deviations for the empirical density of 
excursions between consecutive jumps is described.
These various large deviations properties are then illustrated 
with three examples of positive jump-drift models without diffusion $D(x)=0$ in sections \ref{sec_reset0}, \ref{sec_blowup}, \ref{sec_rain}, and with two examples of jump-diffusion models in sections \ref{sec_diffusionresetR} and \ref{sec_OUavecjumpHz}.
Our conclusions are summarized in \ref{sec_conclusion}.


\section{ Jump-Diffusion Process converging towards some localized steady-state }

\label{sec_diff}

\subsection{ Diffusion with drift $v(x)$ and diffusion coefficient $D(x)$ ; jumps with rate $\lambda(x)$ and probability $\Pi(x' \vert x)$}

We consider the following one-dimensional jump-diffusion dynamics : when at position $x$,
the particle experiences the drift $v(x)$ and diffuses with the diffusion coefficient $D(x)$,
but with the jump rate $\lambda(x)$ per unit time, it can also make a non-local jump toward
 some new position $x'$ chosen with the probability distribution $\Pi(x' \vert x) $
normalized to unity for any $x$
\begin{eqnarray}
\int dx' \Pi(x' \vert x) =1
\label{normaPi}
\end{eqnarray}
As a consequence, the probability $\rho_t(x) $ to be at position $x$ at time $t$ 
satisfies the continuity equation
\begin{eqnarray}
\frac{ \partial \rho_t(x)   }{\partial t}   =  -   \frac{\partial j_t(x) }{\partial x}  -  Q^-_t(x)+  Q^+_t(x)
\label{jumpdiff}
\end{eqnarray}
with the following notations.
The diffusive current $j_t(x) $ involves the drift $v(x)$ and the diffusion coefficient $D(x)$
\begin{eqnarray}
j_t(x) \equiv v(x) \rho_t(x )   -D(x)  \frac{\partial  \rho_t(x) }{\partial x}
\label{jtx}
\end{eqnarray}
The out-flow $Q^-_t(x) $ out of the position $x$ involves the jump rate $\lambda(x)$ 
\begin{eqnarray}
Q^-_t(x)   \equiv  \lambda( x)   \rho_t(x)
\label{qt-}
\end{eqnarray}
The in-flow $Q^+_t(x) $ into the position $x$ takes into account
 the arrival after a jump from any other position $y$,
so that it involves both the jump rate $\lambda(y)$ and the jump probability $\Pi(x \vert y) $
\begin{eqnarray}
Q^+_t(x)   \equiv  \int d y  \  \Pi(x \vert y) \lambda( y) \rho_t(y)
\label{qt+}
\end{eqnarray}
The total probability $n_t$ of a jump at time $t$ can be computed either via integration over the out-flow $Q^-_t(x)   $
or via integration over the in-flow $ Q^+_t(x) $
\begin{eqnarray}
n_t   \equiv \int dx Q^-_t(x)  = \int dx \lambda( x)   \rho_t(x) = \int dx Q^+_t(x) 
\label{nt}
\end{eqnarray}


\subsection{ Existence of a normalizable steady state }

In the whole paper, we will assume that the steady-state solution $\rho_*(x)$ of Eq. \ref{jumpdiff}
\begin{eqnarray}
0 = -  \frac{ d }{ d x}   \left[ v(x) \rho_*(x )  -D (x)   \frac{ d  \rho_*(x)}{ d x}    \right]
-  \lambda( x)   \rho_*(x) +  \int d y  \ \Pi(x \vert y) \lambda( y)  \rho_*(y)
\label{jumpdiffst}
\end{eqnarray}
is normalizable
\begin{eqnarray}
\int dx  \rho_*(x) =1
\label{normast}
\end{eqnarray}
 The corresponding steady current reads
\begin{eqnarray}
j_*(x) && = \rho_*(x) v(x)-D(x)  \rho_*'(x)
\label{jst}
\end{eqnarray}
while the steady out-flow and in-flow are given by
\begin{eqnarray}
Q^-_*(x)  && =  \lambda( x)   \rho_*(x)
\nonumber \\
Q^+_*(x)  && = \int d y  \  \Pi(x \vert y) \lambda( y) \rho_*(y)
\label{qinoutst}
\end{eqnarray}
with the corresponding steady density of jumps 
\begin{eqnarray}
n_*   \equiv \int dx Q^-_*(x)  = \int dx \lambda( x)   \rho_*(x) = \int dx Q^+_*(x) 
\label{nst}
\end{eqnarray}

The goal of the present paper is to analyze the possible fluctuations around these steady state properties.


\section{ Large deviations at level 2.5 for the empirical density and flows  }

\label{sec_2.5}

\subsection{ Empirical density $\rho(x) $, empirical current $j(x)$ and empirical jump-flow $Q(x,y) $ with their constraints}

For a very long dynamical trajectory
$x(0 \leq t \leq T)$ of the jump-diffusion process of Eq. \ref{jumpdiff},
the relevant empirical time-averaged observables are :

(i) the empirical time-averaged density
\begin{eqnarray}
 \rho(x) && \equiv \frac{1}{T} \int_0^T dt \  \delta ( x(t)- x)  
\label{rhodiff}
\end{eqnarray}
normalized to unity
\begin{eqnarray}
\int d x \ \rho (x) && = 1
\label{rho1ptnormadiff}
\end{eqnarray}

(ii) the empirical time-averaged current $j(x)$ characterizing the drift-diffusive part of the dynamics
\begin{eqnarray} 
j(x) \equiv   \frac{1}{T} \int_0^T dt \ \frac{d x(t)}{dt}   \delta( x(t)- x)  
\label{diffjlocal}
\end{eqnarray}

(iii) the empirical time-averaged jump-flow $Q(x,y) $ measuring the density of jumps from $y $ towards $x$
\begin{eqnarray}
Q(x,y)  \equiv  \frac{1}{T} \sum_{t : x(t^+)  \ne  x(t^-) } \delta(x(t^+) - x )  \delta(x(t^-) - y ) 
\label{Qflow}
\end{eqnarray}
The corresponding empirical time-averaged out-flow $Q^-(.) $ and in-flow $Q^+(.) $ 
can be obtained via integration over one position
\begin{eqnarray}
Q^-(y)  \equiv  \frac{1}{T} \sum_{t : x(t^+)  \ne  x(t^-) }  \delta(x(t^-) - y ) =\int dx Q(x,y)  
\nonumber \\
Q^+(x)  \equiv   \frac{1}{T} \sum_{t : x(t^+)  \ne  x(t^-) } \delta(x(t^+) - x )  = \int dy Q(x,y)  
\label{Qflowpm}
\end{eqnarray}
while the total density $n$ of jumps during $[0,T]$ corresponds to the integration over the two positions
\begin{eqnarray}
n \equiv \frac{1}{T} \sum_{t : x(t^+)  \ne  x(t^-) } 1 = \int dx \int dy Q(x,y)= \int dx Q^+( x) = \int dy Q^-( y)  
\label{ndensity}
\end{eqnarray}

For any position $x$, the stationarity constraint involves the divergence of the current $j(x)$ and the difference between the in-flow $Q^+(x) $ and the out-flow $Q^-(x) $ 
\begin{eqnarray}
0 =  - \frac{ dj(x)}{dx}  - Q^-(x)   + Q^+(x)  
\label{statio}
\end{eqnarray}
The integral version of this stationary constraint reads
\begin{eqnarray}
  j( x) = \int_x^{+\infty} dy \left[Q^-( y)  - Q^+( y)  \right] =   \int_{-\infty}^x dy \left[Q^+( y)  - Q^-( y)  \right] 
\label{jpintegral}
\end{eqnarray}
where the vanishing of the full integral is a consequence of Eq. \ref{ndensity}
\begin{eqnarray}
 \int_{-\infty}^{+\infty} dy \left[Q^+( y)  - Q^-( y)  \right] = n-n =0
\label{vanishing}
\end{eqnarray}
Eq. \ref{jpintegral} can also be rewritten in terms of the jump-flow $Q(.,.)$ as
\begin{eqnarray}
  j(z) =     \int_z^{+\infty}  d y \int_{-\infty}^z d x \ Q(x,y)
-  \int_{-\infty}^z dy \int_z^{+\infty} dx \ Q(x,y)
\label{jrecoverQ}
\end{eqnarray}
with the following physical meaning :
the backward jump-flow $Q(x,y) $ with $x<y$ have to be compensated by a positive current contribution at 
any point of the interval $z \in ]x,y[$ (first term),
while the forward jump-flow $Q(x,y) $ with $x>y$ have to be compensated by a negative current contribution at 
any point of the interval $z \in ]y,x[$ (second term).
In particular, the global integral over $z$ of the current $j(z)$ of Eq. \ref{jrecoverQ}
\begin{eqnarray}
\int dz  \  j(z) =  -    \int d y \int d x \ (x-y) Q(x,y)
\label{jrecoverQinteg}
\end{eqnarray}
has to compensate the average amplitude $(x-y)$ of the jump-flow $Q(x,y)$.


\subsection{ Large deviations at Level 2.5 for the empirical time-averaged observables }

The joint distribution of the empirical density $\rho(.)$, of the empirical current $j(.)$
and of the empirical jump-flow $Q(.,.)$, with the corresponding in-flow $Q^+(.)$, out-flow $Q^-(.)$
and density $n$ 
satisfy the large deviation form 
\begin{eqnarray}
 P^{2.5}_T[ \rho(.), j(.),n,Q^{\pm}(.),Q(.,.)]   \opsimeq_{T \to +\infty}  
C_{2.5}[ \rho(.), j(.),n,Q^{\pm}(.),Q(.,.)] 
e^{- \displaystyle T I_{2.5} [ \rho(.), j(.),Q(.,.)]}
\label{ld2.5diff}
\end{eqnarray}
The prefactor $C_{2.5}[ \rho(.), j(.),n,Q^{\pm}(.),Q(.,.)]  $ contains the constitutive constraints 
of normalization (Eq. \ref{rho1ptnormadiff}) and stationarity (Eq. \ref{statio}),
as well as the definitions of the in-flow $Q^+(.) $, of the out-flow $Q^-(.) $ and of the jump density $n$
in terms of the jump flow $Q(.,.)$ (Eqs \ref{Qflowpm} and \ref{ndensity})
\begin{eqnarray}
&& C_{2.5}[ \rho(.), j(.),n,Q^{\pm}(.),Q(.,.)]   = \delta \left(\int d x \rho(x) -1  \right)
\left[ \prod_{x }  \delta \left(  j'(x) + Q^-(x)   -  Q^+(x)  \right) \right]     
 \label{c2.5}      \\ && 
 \delta \left(  \int d y \ Q^-(y) - n     \right)    \delta \left(  \int d x \ Q^+(x) - n     \right)     
 \left[ \prod_{ x } \delta \left(  \int d y \  Q(x,y) - Q^+(x)     \right)   \right]
\left[ \prod_{ y } \delta \left(  \int d x \ Q(x,y) - Q^-(y) \right) \right]
 \nonumber
\end{eqnarray}
The rate function $I_{2.5} [ \rho(.), j(.),Q(.,.)]$ contains two contributions 
\begin{eqnarray}
 I_{2.5} [ \rho(.), j(.),Q(.,.)]= I_{2.5}^{[D,v]} [ \rho(.), j(.)]   +  I_{2.5}^{[\lambda,\Pi]} [ \rho(.), Q(.,.)]
\label{rate2.5}
\end{eqnarray}

(i) The first contribution involving the diffusion coefficient $D(x)$ and the drift $v(x)  $
corresponds to the usual rate function for diffusion processes \cite{wynants_thesis,maes_diffusion,chetrite_formal,engel,chetrite_HDR,c_reset,c_lyapunov,c_inference}
\begin{eqnarray}
 I_{2.5}^{[D,v]} [ \rho(.), j(.)]  
 = 
\int \frac{d x}{ 4 D(x) \rho(x) } \left[ j(x) - \rho(x) v(x)+D(x) \rho' (x) \right]^2
\label{rate2.5diff}
\end{eqnarray}

(ii)The second contribution involving the jump rate $\lambda(y)$ and the jump probability $\Pi(x \vert y)$
corresponds to the usual rate function for Markov jump processes \cite{fortelle_thesis,fortelle_jump,maes_canonical,maes_onandbeyond,wynants_thesis,chetrite_formal,BFG1,BFG2,chetrite_HDR,c_ring,c_interactions,c_open,c_detailed,barato_periodic,chetrite_periodic,c_reset,c_inference}
\begin{eqnarray}
 I_{2.5}^{[\lambda,\Pi]} [ \rho(.), Q(.,.)] =  \int d x \int d y
\left[ Q(x,y) \ln \left( \frac{ Q(x,y)  }{   \Pi(x \vert y) \lambda( y)  \rho(y) }  \right) 
 - Q(x,y) +  \Pi(x \vert y) \lambda( y)  \rho(y)  \right]
\label{rate2.5jump}
\end{eqnarray}
Using the normalization of Eq. \ref{normaPi} and the constraints of Eq. \ref{c2.5},
it is useful to introduce the out-flow $Q^-(y)$ in order to rewrite
this jump rate function as a sum of two terms
\begin{eqnarray}
 I^{[\lambda,\Pi]}_{2.5} [ \rho(.), Q^-(.),Q(.,.)]   =  I^{[\lambda]}_{2.5} [ \rho(.), Q^-(.)]
 +  I^{[\Pi]}_{2.5} [  Q^-(.),Q(.,.)]
\label{rate2.5jumplambdapi}
\end{eqnarray}
The first term associated to the jump rate $\lambda(.)$
\begin{eqnarray}
 I^{[\lambda]}_{2.5} [ \rho(.), Q^-(.)]   = \int d y \left[ Q^-(y) \ln \left( \frac{  Q^-(y)  }{    \lambda(y)   \rho(y) }  \right) 
 -  Q^-(y) +   \lambda(y)   \rho(y)  \right]
\label{rate2.5jumplambda}
\end{eqnarray}
describes the possible fluctuations of the out-flow $Q^-(y)$ with respect to its typical value $\lambda(y)   \rho(y) $.
The second term associated to the jump probability $\Pi(.\vert.)$
\begin{eqnarray}
 I^{[\Pi]}_{2.5} [  Q^-(.),Q(.,.)]  = 
 \int d x \int d y Q(x,y) \ln \left( \frac{ Q(x,y)  }{   \Pi(x \vert y) Q^-(y) }  \right) 
\label{rate2.5jumppi}
\end{eqnarray}
characterizes the possible fluctuations of the empirical jump probability $\frac{ Q(x,y)  }{ Q^-(y) }$
with respect to its typical value $\Pi( x \vert y)  $.

In summary, the rate function at Level 2.5 of Eq. \ref{rate2.5}
has been decomposed as the sum of the three terms 
associated to the diffusion parameters $[D(.),v(.)]$, to the jump rate $\lambda(.)$
and to the jump probability $\Pi(.\vert.)$ respectively
\begin{eqnarray}
 I_{2.5} [ \rho(.), j(.),Q^-(.),Q(.,.)]= I_{2.5}^{[D,v]} [ \rho(.), j(.)]  
  +  I^{[\lambda]}_{2.5} [ \rho(.), Q^-(.)]  
  + I^{[\Pi]}_{2.5} [  Q^-(.),Q(.,.)] 
\label{rate2.5trois}
\end{eqnarray}


\subsection{ Alternative formulation for the inferred parameters that would make the empirical observables typical}

Another point of view on the large deviations at Level 2.5 is based 
on the inverse problem of inference \cite{c_inference} :
from the data of a long dynamical trajectory, 
one computes the empirical time-averaged observables described above,
and one infers the best steady state $\hat \rho_*(x) $
and the best corresponding parameters $[\hat v(x), \hat \lambda(x), \hat \Pi(x \vert y)]$ of the model as follows
(Note that the diffusion coefficient cannot fluctuate $\hat D(x) \equiv D(x)$ as discussed in detail in \cite{c_inference}).

(i) the best inferred steady state $\hat \rho_*(x) $ is simply the measured empirical density 
\begin{eqnarray}
\hat \rho_* (x) \equiv \rho(x)
\label{hatrho}
\end{eqnarray}

(ii) the best inferred drift $ \hat v(x)$ is the drift that would make vanish the diffusive rate function Eq. \ref{rate2.5diff}
\begin{eqnarray}
\hat v(x) \equiv \frac{j(x) +D(x) \rho' (x) }{\rho(x) }
\label{hatv}
\end{eqnarray} 

(iii) the best inferred jump rate $\hat \lambda(y) $ is the rate that would make vanish the rate function of Eq. \ref{rate2.5jumplambda}
\begin{eqnarray}
\hat \lambda(y) \equiv  \frac{  Q^-(y)  }{   \rho(y) } 
\label{hatlambda}
\end{eqnarray}

(iv) the best inferred jump probability $\hat \Pi(x \vert y) $ is the jump probability that would make vanish the rate function of Eq. \ref{rate2.5jumppi}
\begin{eqnarray}
 \hat \Pi(x \vert y) \equiv \frac{ Q(x,y)  }{  Q^-(y) }  
\label{hatpi}
\end{eqnarray}

Via this change of variables, the large deviations at Level 2.5 of Eq. \ref{ld2.5diff} translates into 
the joint probability to infer the model parameters $[\hat v(x), \hat \lambda(x), \hat \Pi(x \vert y)]$
and the corresponding steady state $\hat \rho_*(x) $ that they produce together
\begin{eqnarray}
 P^{Infer}_T[ \hat \rho_* (.),\hat v(.), \hat \lambda(.), \hat \Pi(. \vert .) ]   \opsimeq_{T \to +\infty}  
C_{Infer} [ \hat \rho_* (.),\hat v(.), \hat \lambda(.), \hat \Pi(. \vert .) ] 
e^{- \displaystyle T I_{Infer} [ \hat \rho_* (.),\hat v(.), \hat \lambda(.), \hat \Pi(. \vert .) ] }
\label{ldinfer}
\end{eqnarray}
The constraints derived from Eq. \ref{c2.5}
\begin{eqnarray}
&& C_{Infer}[\hat \rho_* (.),\hat v(.), \hat \lambda(.), \hat \Pi(. \vert .) ]   = \delta \left(\int d x \hat \rho_*(x) -1  \right)
\left[ \prod_{ y } \delta \left(  \int d x \  \hat \Pi(x \vert y) - 1 \right) \right]
 \nonumber \\
&& \left[ \prod_{x }  \delta \left( \frac{d}{dx} \left[ \hat \rho_*(x) \hat v(x)-D(x) \frac{d \hat \rho_* (x)}{dx} \right] 
 + \hat \lambda(x)  \hat \rho_*(x)  -  \int d y \  \hat \Pi(x \vert y) \hat \lambda(y)  \hat \rho_*(y) \right) \right]     
 \label{cinfer}      
\end{eqnarray}
contains the normalization of the inferred steady state $ \hat \rho_* (.)$ and the normalization of the inferred jump probability 
$\hat \Pi(. \vert .) $ on the first line, while the second line means that $\hat \rho_* (.) $ 
should be the steady state produced by the inferred parameters $\hat v(.), \hat \lambda(.), \hat \Pi(. \vert .) $.
The rate function translated from Eq. \ref{rate2.5trois} contains three contributions
\begin{eqnarray}
 I_{Infer} [ \hat \rho_* (.),\hat v(.), \hat \lambda(.), \hat \Pi(. \vert .) ] && =
   I_{Infer}^{[D,v]} [ \hat \rho_*(.), \hat v (.)]  
   +  I^{[\lambda]}_{Infer} [\hat \rho_*(.), \hat \lambda(.) ]
   +   I^{[\Pi]}_{Infer} [\hat \rho_*(.), \hat \lambda(.),  \hat \Pi(. \vert .) ] 
\label{rateinfer}
\end{eqnarray}
The first contribution governs the fluctuations of the inferred drift $\hat v (.) $ around the true drift $v(.)$
\begin{eqnarray}
  I_{Infer}^{[D,v]} [ \hat \rho_*(.), \hat v (.)]  && =
\int  dx \frac{ \hat \rho_* (x)}{ 4 D(x)  } \left[ \hat v(x) - v(x) \right]^2
\label{rateinfer1}
\end{eqnarray}
The second contribution governs the fluctuations of the inferred jump rate $\hat \lambda(.) $
around the true jump rate $ \lambda(.) $
\begin{eqnarray}
    I^{[\lambda]}_{Infer} [\hat \rho_*(.), \hat \lambda(.) ]  && = \int d y \hat \rho_*(y)
    \left[ \hat \lambda(y)   \ln \left( \frac{   \hat \lambda(y)   }{    \lambda(y)    }  \right) 
 - \hat \lambda(y)  +   \lambda(y)    \right]
\label{rateinfer2}
\end{eqnarray}
The third contribution governs the fluctuations of the inferred jump probability $\hat  \Pi(. \vert .) $
with respect to the true jump probability $  \Pi(. \vert .) $
\begin{eqnarray}
   I^{[\Pi]}_{Infer} [\hat \rho_*(.), \hat \lambda(.),  \hat \Pi(. \vert .) ]  && =
   \int d y  \hat \rho_*(y) \hat \lambda(y) 
   \int d x   \hat \Pi(x \vert y) 
    \ln \left( \frac{  \hat \Pi(x \vert y)  }{   \Pi(x \vert y) }  \right) 
\label{rateinfer3}
\end{eqnarray}


\subsection{ Simplification of the Level 2.5 for jump-drift models without diffusion $D(x)=0$}

For jump-drift models without diffusion $D(x)=0$, the diffusive rate function of Eq. \ref{rate2.5diff}
does not appear anymore
but yields that the empirical current is directly related to the empirical density
\begin{eqnarray}
j(x)  =  \rho(x) v(x) \ \ {\rm when } \ \ D(x)=0
\label{jrhonodiff}
\end{eqnarray} 
as a consequence of the deterministic motion at velocity $v(x)$ between jumps.

So the large deviations at Level 2.5 of Eq. \ref{ld2.5diff} can be rewritten
after the elimination of the current $j(.)$ via Eq. \ref{jrhonodiff}
\begin{eqnarray}
 P^{2.5[D(.)=0]}_T[ \rho(.), n,Q^{\pm}(.),Q(.,.)]   \opsimeq_{T \to +\infty}  
C^{[D(.)=0]}_{2.5}[ \rho(.), n,Q^{\pm}(.),Q(.,.)] 
e^{- \displaystyle T  I_{2.5}^{[D(.)=0]} [ \rho(.), Q^{-}(.),Q(.,.)]}
\label{ld2.5nodiff}
\end{eqnarray}
with the constraints
\begin{eqnarray}
&& C^{[D(.)=0]}_{2.5}[ \rho(.), n,Q^{\pm}(.),Q(.,.)]   =   
 \delta \left(\int d x \rho(x) -1  \right)
\left[  \prod_{x } \delta \left(  \frac{d}{dx} \left[  \rho(x) v(x)\right] + Q^-(x)   -  Q^+(x)  \right) \right]     
 \label{c2.5nodiff}      \\ && 
 \delta \left(  \int d y \ Q^-(y) - n     \right)    \delta \left(  \int d x \ Q^+(x) - n     \right)     
 \left[ \prod_{ x } \delta \left(  \int d y \  Q(x,y) - Q^+(x)     \right)   \right]
\left[ \prod_{ y } \delta \left(  \int d x \ Q(x,y) - Q^-(y) \right) \right]  
 \nonumber   
 \end{eqnarray}
 and the rate function involving only the two jump contributions associated to $\lambda(.)$ and $\Pi(.\vert.)$
 \begin{eqnarray}
 I_{2.5}^{[D(.)=0]}&& \! \! \! \! \! \! [ \rho(.), Q^{-}(.),Q(.,.)]  =  I^{[\lambda]}_{2.5} [ \rho(.), Q^-(.)]  
  + I^{[\Pi]}_{2.5} [  Q^-(.),Q(.,.)] 
    \nonumber \\
  && =\int d y \left[ Q^-(y) \ln \left( \frac{  Q^-(y)  }{    \lambda(y)   \rho(y) }  \right) 
 -  Q^-(y) +   \lambda(y)   \rho(y)  \right]
 +\int d x \int d y Q(x,y) \ln \left( \frac{ Q(x,y)  }{   \Pi(x \vert y) Q^-(y) }  \right) 
\label{rate2.5nodiff}
\end{eqnarray}
 Three examples of jump-drift models without diffusion $D(x)=0$ 
 will be described in sections \ref{sec_reset0}, \ref{sec_blowup}, \ref{sec_rain}.


\subsection{ Simplification of the Level 2.5 when the jump probability $\Pi^{deter}( x \vert y)  =  \delta(x-\Phi(y))$ is deterministic }

When the jump probability describes a deterministic rule for the position after the jump $x=\Phi(y)$
in terms of the position $y$ before the jump
\begin{eqnarray}
\Pi^{deter}( x \vert y) =  \delta \left(x - \Phi(y) \right) 
 \label{jumpdeter}
\end{eqnarray}
 the jump flow involves the same deterministic function $\delta \left(x - \Phi(y) \right)  $
\begin{eqnarray}
 Q(x , y) =  \delta \left(x - \Phi(y) \right) Q^-(y)
 \label{jumpflowdeter}
\end{eqnarray}
and the rate function of Eq. \ref{rate2.5jumppi} vanishes
\begin{eqnarray}
 I^{[\Pi^{deter}]}_{2.5} [  Q^-(.),Q(.,.)]  = 0
\label{rate2.5jumppideter}
\end{eqnarray}
One can also use Eq. \ref{jumpflowdeter}
to eliminate the in-flow $Q^+(.)$ in terms of the out-flow $Q^-(.)$
\begin{eqnarray}
 Q^+(x) =  \int d y Q(x , y)  = \int d y \  \delta \left(x - \Phi(y) \right)  Q^-(y) 
\label{inflowdeter}
\end{eqnarray}
Putting everything together, one obtains that the large deviations at Level 2.5 of Eq. \ref{ld2.5diff}
can be rewritten for the joint distribution of the density $\rho(.)$, of the current $j(.)$,
of the density $n$ and of the out-flow $Q^-(.)$ only as
\begin{eqnarray}
&& P_T^{2.5[\Pi^{deter}( x \vert y) =  \delta \left(x - \Phi(y) \right) ]}[ \rho(.), j(.),n,Q^{-}(.)]   \opsimeq_{T \to +\infty}  
 \delta \left(\int d x \rho(x) -1  \right) 
 \delta \left(  \int d y \ Q^-(y) - n     \right)   
\nonumber \\
&&
\left[ \prod_{x }  \delta \left(  j'(x) + Q^-(x)   -   \int d y \  \delta \left(x - \Phi(y) \right)  Q^-(y)   \right) \right]    
e^{- \displaystyle T  \left[ I_{2.5}^{[D,v]} [ \rho(.), j(.)]+ I^{[\lambda]}_{2.5} [ \rho(.), Q^-(.)] \right]}
\label{ld2.5diffpideter}
\end{eqnarray}
Two examples of deterministic jump probability $\Pi^{deter}( x \vert y)  =  \delta(x-\Phi(y))$
will be described in sections \ref{sec_reset0} and \ref{sec_blowup}.


\section{ Contraction over the jump-flow $Q(x,y) $ for given in/out-flows $Q^{\pm}(.)$ }

\label{sec_contractionQflow}

In the large deviations at Level 2.5 described in the previous section, the jump-flow $Q(x,y) $
is the only empirical observable that involves two positions.
The goal of this section is to analyze whether the contraction of the jump-flow $Q(x,y) $
can be carried out.


\subsection{ Large deviations for the one-position empirical observables $[ \rho(.), j(.),Q^{\pm}(.)] $ 
via contraction over $Q(.,.)$}

The joint distribution of $\rho(.)$, $j(.)$, $n$ and $Q^{\pm}(.)$
can be derived from the Level 2.5 of Eq. \ref{ld2.5diff}
 \begin{eqnarray}
&& P_T[ \rho(.), j(.),n,Q^{\pm}(.)] 
    \opsimeq_{T \to +\infty}  
 \delta \left(\int d x \rho(x) -1  \right)
\left[ \prod_{x }  \delta \left(  j'(x) + Q^-(x)   -  Q^+(x)  \right) \right]     
      \\ && 
 \delta \left(  \int d y \ Q^-(y) - n     \right)    \delta \left(  \int d x \ Q^+(x) - n     \right)     
 e^{- \displaystyle T \left( 
I_{2.5}^{[D,v]} [ \rho(.), j(.)]  
+I^{[\lambda]}_{2.5} [ \rho(.), Q^-(.)] 
\right) } K_T
\label{ld2.5diffpicontraction}
\end{eqnarray}
if one can evaluate the remaining integral over the jump flow $Q( x , y) $
\begin{eqnarray}
K_T && \equiv \int {\cal D} Q(.,.) 
 \left[ \prod_{ x } \delta \left(  \int d y \  Q(x,y) - Q^+(x)     \right)   \right]
\left[ \prod_{ y } \delta \left(  \int d x \ Q(x,y) - Q^-(y) \right) \right]
\nonumber \\ && 
e^{- \displaystyle T \int d x \int d y Q(x,y) \ln \left( \frac{ Q(x,y)  }{   \Pi(x \vert y) Q^-(y) }  \right)   }
\label{kintegral}
\end{eqnarray}
for large $T$.
In order to optimize over the jump flow $Q( . , .) $
the functional appearing in factor of $T$ in the exponential of the second line
in the presence of the constraints of the first line,
let us introduce the following Lagrangian 
 \begin{eqnarray}
 {\cal L}  [Q( . , .) ] 
&& \equiv  - \int d x \int d y Q(x,y) \ln \left( \frac{ Q(x,y)  }{   \Pi(x \vert y) Q^-(y) }  \right)   
 \nonumber \\ &&
 +\int dx \phi(x) \left( \int d y \  Q(x,y) - Q^+(x)    \right)  
+ \int dy  \psi(y) \left( \int d x \ Q(x,y) - Q^-(y)  \right)   
 \label{lagrangeqq}
\end{eqnarray}
where the Lagrange multipliers $ \phi(.),\psi(.)$ are associated to the two constraints.
The optimization with respect to the jump flow $Q( x , y)$ 
 \begin{eqnarray}
0  = \frac{ \partial {\cal L}[Q( . , .)]  }{\partial Q( x, y) }
 =      -  \ln \left( \frac{ Q(x,y)  }{   \Pi(x \vert y) Q^-(y) }  \right)    -1
 + \phi(x)   
+   \psi(y) 
 \label{lagrangeqqderi}
\end{eqnarray}
leads to the optimal solution
 \begin{eqnarray}
Q^{opt}(x,y)   = e^{-1}  e^{  \phi(x)} \Pi(x \vert y) Q^-(y) e^{  \psi(y) }
\label{qqopt}
\end{eqnarray}
that should satisfy the two constraints 
\begin{eqnarray}
Q^+(x)  && =   \int dy  Q^{opt}(x,y)  = e^{-1}  e^{  \phi(x)} \int dy \Pi(x \vert y) Q^-(y) e^{  \psi(y) }
\nonumber \\
Q^-(y)  && =   \int dx  Q^{opt}(x,y) =  e^{-1} Q^-(y) e^{  \psi(y) } \int dx e^{  \phi(x)}  \Pi(x \vert y) 
\label{jumpflowsrecoveropt}
\end{eqnarray}
The second constraint 
can be used to eliminate the Lagrange multiplier $\psi(y)$ in terms of the other one $\phi(x)$
\begin{eqnarray}
  e^{\psi(y)}  =\frac{1}{ e^{-1} \int dx' e^{  \phi(x')  } \Pi(x' \vert y) }
\label{psisol}
\end{eqnarray}
Plugging this value into Eq. \ref{qqopt} yields
 \begin{eqnarray}
Q^{opt}(x,y)   =    \frac{e^{  \phi(x)} \Pi(x \vert y)}{  \int dx' e^{  \phi(x')  } \Pi(x' \vert y) }  Q^-(y)
\label{lagrangeqqopt}
\end{eqnarray}
while the first constraint of Eq. \ref{jumpflowsrecoveropt} becomes
\begin{eqnarray}
Q^+(x)   =  e^{  \phi(x)}  \int dy  \frac{ \Pi(x \vert y)}{  \int dx' e^{  \phi(x')  } \Pi(x' \vert y) }  Q^-(y)
\label{phisol}
\end{eqnarray}
Let us now describe two examples where this optimization problem has a simple solution.


\subsection{ Explicit contraction for resetting models $\Pi^{reset}( x \vert y)  =  R(x) $  }

When the jump probability is independent of the starting point $y$
\begin{eqnarray}
\Pi^{reset}( x \vert y)  =  R(x) 
\label{reset}
\end{eqnarray}
the jumps correspond to the following stochastic resetting procedure
 (see the review \cite{review_reset} and references therein) :
 $\lambda(y)$ is the reset rate when in position $y$, while 
the normalized probability distribution $R(x)$ governs the choice of the new position $x$ after each jump.
Then Eq. \ref{phisol} reduces to
\begin{eqnarray}
Q^+(x)   =    \frac{ e^{  \phi(x)}  R(x)}{  \int dx' e^{  \phi(x')  } R(x') } \int dy Q^-(y)
=  \frac{ e^{  \phi(x)}  R(x)}{  \int dx' e^{  \phi(x')  } R(x') } n
\label{phisolreset}
\end{eqnarray}
and the optimal solution $Q^{opt}(x,y) $ of Eq. \ref{lagrangeqqopt} 
can be rewritten in terms of the in-flow $Q^+(x)$ at $x$, of the out-flow $Q^-(y)$ at the position $y$ and 
of the jump density $n$ 
 \begin{eqnarray}
Q^{opt}(x,y)   =    \frac{e^{  \phi(x)} R(x)}{  \int dx' e^{  \phi(x')  } R(x') }  Q^-(y)  = \frac{Q^+(x) Q^-(y)}{n}
\label{lagrangeqqderioptreset}
\end{eqnarray}
The value of the Lagrangian of Eq. \ref{lagrangeqq} for this optimal solution
satisfying the constraints reduces to
 \begin{eqnarray}
 {\cal L}  [Q^{opt}(.,.)] 
&& \equiv 
  - \int d x \int d y Q^{opt}(x,y) \ln \left( \frac{ Q^{opt}(x,y)  }{   R(x) Q^-(y) }  \right)   
 =   - \int d x \int d y \frac{Q^+(x) Q^-(y)}{n} \ln \left( \frac{ Q^+(x)   }{  n R(x)  }  \right)   
  \nonumber \\
&&
 = -    \int d x  Q^+(x) \ln \left( \frac{Q^+(x)  }{ n R(x) } \right) \equiv - I^{[R]} [ n, Q^+(.)]
 \label{lagrangeqqcalcul}
\end{eqnarray}
where the rate function
 \begin{eqnarray}
I^{[R]} [ n, Q^+(.)] \equiv    \int d x  Q^+(x) \ln \left( \frac{Q^+(x)  }{ n R(x) } \right) 
 \label{ratereset}
\end{eqnarray}
governs the possible fluctuations of the out-flow $Q^+(x)$ with respect to its typical value $n R(x) $
involving the normalized reset probability $R(x)$.
The optimal value of Eq. \ref{lagrangeqqcalcul}
governs the exponential behavior in $T$ of the integral of Eq. \ref{kintegral}
 \begin{eqnarray}
 K_T  \opsimeq_{T \to +\infty} e^{ \displaystyle T {\cal L}[\pi^{opt}(.\vert.)]  } 
  = e^{  \displaystyle - T I^{[R]} [ n, Q^+(.)]    } 
 = e^{  \displaystyle - T     \int d x  Q^+(x) \ln \left( \frac{Q^+(x)  }{ n R(x) } \right) } 
\label{integralsaddlefin}
\end{eqnarray}
Plugging this result into Eq. \ref{ld2.5diffpicontraction} yields the large deviation form
for the one-position empirical observables $[ \rho(.), j(.),n,Q^{\pm}(.)]  $
\begin{eqnarray}
&& P_T^{\Pi^{reset}( x \vert y) =R(x)}[ \rho(.), j(.),n,Q^{\pm}(.)] 
    \opsimeq_{T \to +\infty}  
 \delta \left(\int d x \rho(x) -1  \right)
\left[ \prod_{x }  \delta \left(  j'(x) + Q^-(x)   -  Q^+(x)  \right) \right]     
\label{ld2.5diffpicontractionfinal}      \\ && 
 \delta \left(  \int d y \ Q^-(y) - n     \right)    \delta \left(  \int d x \ Q^+(x) - n     \right)     
 e^{- \displaystyle T \left( 
I_{2.5}^{[D,v]} [ \rho(.), j(.)]  
+I^{[\lambda]}_{2.5} [ \rho(.), Q^-(.)] 
+  I^{[R]} [ n, Q^+(.)] 
\right) }
\nonumber 
\end{eqnarray}
An example of jump-diffusion process with the resetting jump probability 
$\Pi^{reset}( x \vert y)  =  R(x)$ towards an aritrary function $R(x)$ 
will be described in section \ref{sec_diffusionresetR}.


\subsection{ Explicit contraction for the positive exponential jump probability $\Pi^{exp}_{\alpha}(x \vert y)= \alpha e^{- \alpha (x-y) }$ for $x \geq y$ }

Let us now consider the case where the jump probability 
\begin{eqnarray}
 \Pi^{exp}_{\alpha}(x \vert y)= \alpha e^{- \alpha (x-y) } \ \ {\rm for } \ \ x \in [y,+\infty[
\label{exp}
\end{eqnarray}
describes positive jumps whose amplitude $z=x-y \geq 0 $ is exponentially distributed.
Then Eq. \ref{lagrangeqqopt} reads for $x \geq y$
 \begin{eqnarray}
Q^{opt}(x,y)   =    \frac{e^{  \phi(x)}   e^{- \alpha x }}
{  \int_y^{+\infty}  dx''e^{  \phi(x'')  }   e^{- \alpha x'' } }  Q^-(y)
\label{lagrangeqqderioptexp}
\end{eqnarray}
while Eq. \ref{phisol} becomes
\begin{eqnarray}
Q^+(x)   =  \int_{-\infty}^x dy  Q^{opt}(x,y)   = e^{  \phi(x)}   e^{- \alpha x }  \int_{-\infty}^x dy    \frac{Q^-(y)}
{  \int_y^{+\infty}  dx'' e^{  \phi(x'')  }   e^{- \alpha x'' } }  
\label{phisolexp}
\end{eqnarray}
It is useful to introduce the corresponding negative current via Eq. \ref{jrecoverQ}
\begin{eqnarray}
  j(x) && =   -   \int_x^{+\infty} dx' \ \int_{-\infty}^x dy' Q^{opt}(x',y') =   
  -  \left[ \int_x^{+\infty} dx' e^{  \phi(x')}   e^{- \alpha x' } \right] \ 
\left[   \int_{-\infty}^x dy    \frac{ Q^-(y)}{  \int_y^{+\infty}  dx''e^{  \phi(x'')  }   e^{- \alpha x'' } }  \right]
  \label{jrecoverQexp}
\end{eqnarray}
and to consider the ratio
\begin{eqnarray}
\frac{ Q^+(x) } { [- j(x) ] }  =  \frac{ e^{  \phi(x)}   e^{- \alpha x }  } {\int_x^{+\infty} dx' e^{  \phi(x')}   e^{- \alpha x' } } 
= - \frac{d}{dx} \ln \left[ \int_x^{+\infty} dx' e^{  \phi(x')}   e^{- \alpha x' } \right]
\label{ratioqj}
\end{eqnarray}
The integration 
\begin{eqnarray}
\int_y^x dz \frac{ Q^+(z) } { [- j(z) ] }  = \ln \left[ 
\frac{ \int_y^{+\infty} dx' e^{  \phi(x')}   e^{- \alpha x' } }{\int_x^{+\infty} dx' e^{  \phi(x')}   e^{- \alpha x' } }\right]
\label{phisolexpinteg}
\end{eqnarray}
yields
\begin{eqnarray}
 \int_y^{+\infty} dx' e^{  \phi(x')}   e^{- \alpha x' } =
e^{ \int_y^x dz \frac{ Q^+(z) } { [- j(z) ] }  } \int_x^{+\infty} dx' e^{  \phi(x')}   e^{- \alpha x' }  
\label{phisolexpintegy}
\end{eqnarray}
Using this integral and Eq. \ref{ratioqj}, one can rewrite
the optimal solution of Eq. \ref{lagrangeqqderioptexp} 
without the Lagrange multiplier $\phi(.)$ as
 \begin{eqnarray}
Q^{opt}(x,y) &&   =    \frac{e^{  \phi(x)}   e^{- \alpha x }}
{  \int_x^{+\infty}  dx''e^{  \phi(x'')  }   e^{- \alpha x'' } }  
e^{ - \int_y^x dz \frac{ Q^+(z) } { [- j(z) ] }  }
Q^-(y)
= \frac{ Q^+(x) } { [- j(x) ] } e^{ - \int_y^x dz \frac{ Q^+(z) } { [- j(z) ] }  }Q^-(y)
\label{lagrangeqqderioptexpj}
\end{eqnarray}
in terms of the in-flow $Q^+(.)$, of the out-flow $Q^{+}(.)$, and the current $j(.)$.
The value of the Lagrangian of Eq. \ref{lagrangeqq} for this optimal solution
satisfying the constraints reads
 \begin{eqnarray}
 {\cal L}  [Q^{opt}(.,.)] 
&& \equiv 
  - \int d x \int_{-\infty}^x d y \ Q^{opt}(x,y) \ln \left( \frac{ Q^{opt}(x,y)  }{ \alpha e^{-\alpha (x-y) }  Q^-(y) }  \right)   
  \nonumber \\
  && =
  - \int d x \int_{-\infty}^x d y \ Q^{opt}(x,y) \ln \left( \frac{ \frac{ Q^+(x) } { [- j(x) ] } e^{ - \int_y^x dz \frac{ Q^+(z) } { [- j(z) ] }  }  }{ \alpha e^{-\alpha (x-y) }   }  \right)  
    \nonumber \\
  && =
  - \int d x \int_{-\infty}^x d y \ Q^{opt}(x,y) \ln \left( \frac{  Q^+(x)    }{ \alpha   [- j(x) ]  }  \right)  
  - \alpha  \int d x \int_{-\infty}^x d y \ Q^{opt}(x,y)  \left( x-y \right)  
  \nonumber \\  && 
    +\int_{-\infty}^{+\infty} dz \frac{ Q^+(z) } { [- j(z) ] }  \int_z^{+\infty} d x \int_{-\infty}^z d y \ Q^{opt}(x,y)    
     \label{lagrangeqqcalculexp}
\end{eqnarray}
Using the constraints, Eq. \ref{jrecoverQ}  and Eq. \ref{jrecoverQinteg},
Eq. \ref{lagrangeqqcalculexp} reduces to
 \begin{eqnarray}
 {\cal L}  [Q^{opt}(.,.)] 
&& =
  - \int d x  \ Q^{+}(x) \ln \left( \frac{  Q^+(x)    }{ \alpha   [- j(x) ]  }  \right)  
  - \alpha  \int dz  [- j(z) ]
    +\int_{-\infty}^{+\infty} dz  Q^+(z)  
    \nonumber \\
    && = - \int d x  \left[  Q^{+}(x) \ln \left( \frac{  Q^+(x)    }{ \alpha   [- j(x) ]  }  \right)  
    - Q_+(x) + \alpha   [- j(x) ]
    \right] \equiv - I^{exp}_{\alpha} [ Q^+(.), j(.) ]
     \label{lagrangeqqcalculexpfin}
\end{eqnarray}
where the rate function 
 \begin{eqnarray}
I^{exp}_{\alpha} [ Q^+(.), j(.) ] \equiv    \int d x  \left[  Q^{+}(x) \ln \left( \frac{  Q^+(x)    }{ \alpha   [- j(x) ]  }  \right)  
    - Q_+(x) + \alpha   [- j(x) ]
    \right]
 \label{rateexpo}
\end{eqnarray}
governs the possible fluctuations of the out-flow $Q^+(x)$.
The final result is thus that
the large deviation form
for the one-position empirical observables $[ \rho(.), j(.),n,Q^{\pm}(.)]  $ reads
\begin{eqnarray}
&& P_T^{[\Pi^{exp}_{\alpha}(x \vert y)= \alpha e^{- \alpha (x-y) }]}[ \rho(.), j(.),n,Q^{\pm}(.)] 
    \opsimeq_{T \to +\infty}  
 \delta \left(\int d x \rho(x) -1  \right)
\left[ \prod_{x }  \delta \left(  j'(x) + Q^-(x)   -  Q^+(x)  \right) \right]     
\label{ld2.5diffpicontractionfinalexp}      \\ && 
 \delta \left(  \int d y \ Q^-(y) - n     \right)    \delta \left(  \int d x \ Q^+(x) - n     \right)     
 e^{- \displaystyle T \left( 
I_{2.5}^{[D,v]} [ \rho(.), j(.)]  
+I^{[\lambda]}_{2.5} [ \rho(.), Q^-(.)] 
+ I^{exp}_{\alpha} [ Q^+(.), j(.) ] 
\right) }
\nonumber 
\end{eqnarray}
An example of jump-drift process with the exponential jump probability 
$ \Pi^{exp}_{\alpha}(x \vert y)= \alpha e^{- \alpha (x-y) }$ will be described in section \ref{sec_rain}.


\section{ Analysis of the dynamics from the point of view of jump events }

\label{sec_jumpsonly}

In this section, the jump-diffusion dynamics of Eq. \ref{jumpdiff}
is analyzed instead from the point of view of the jump events only.

\subsection{ Alternate Markov chain governing the sequence of jump events}

A very long trajectory $x(0 \leq t \leq T)$ is characterized by a large number $N= n T$ of jumps,
where the density $n$ of jumps had been already discussed in Eq. \ref{ndensity}.
Let us introduce
 the times $t_i$ with $i=1,..,N$ where the $N$ jumps occur, between the positions just before the jump at $t=t_i^-$
 \begin{eqnarray}
y_i \equiv x(t_i^-)
\label{yibefore}
\end{eqnarray}
and the positions just after the jump at $t=t_i^+$
 \begin{eqnarray}
x_i \equiv x(t_i^+)
\label{xiafter}
\end{eqnarray}
The dynamics of these jump events is governed by the following alternate Markov chain 
\begin{eqnarray}
P_{t_{i}^+}(x_i)  && = \int d y_i \ \Pi( x_i \vert y_i) P_{t_{i}^-}(y_i) 
\nonumber \\
P_{t_{i+1}^-}( y_{i+1}) && =  \int d x_i  \ W(t_{i+1}-t_i ,  y_{i+1} \vert x_i) P_{t_{i}^+}(x_i) 
\label{alternatemarkov}
\end{eqnarray}
where the jump kernel $\Pi( x \vert y) $ is given by the definition of the model with the normalization of Eq. \ref{normaPi},
while the kernel $W( \tau , y  \vert x  ) $ for the excursions between two consecutive jumps 
with the normalization
\begin{eqnarray}
  \int_{0}^{+\infty} d \tau \int d y \ W(\tau , y  \vert x  ) =1
\label{normaWesc}
\end{eqnarray}
is discussed in detail below.


\subsection{ Kernel $W(\tau ,  y  \vert x  ) $ for the excursions between two consecutive jumps}

 The probability that an excursion starting at position $x$ ends after the time $\tau$
at the position $y$  
involves the jump rate $ \lambda( y)  $ at position $y$
\begin{eqnarray}
W( \tau, y  \vert x  )
 = \lambda( y)  P^{surv}_{\tau}(  y  \vert x  )
\label{Wesc}
\end{eqnarray}
while $P^{surv}_{\tau}(  y  \vert x  ) $
is the probability to have diffused from the position $x$ to the position $y$ in the time $\tau$ without any jump.
As a consequence, this probability satisfies the initial dynamical Eq. \ref{jumpdiff} with absorption only, i.e. without the last term representing the re-injection after the jumps
\begin{eqnarray}
\frac{ \partial P^{surv}_{\tau}(y \vert x)   }{\partial \tau}   = 
 -   \frac{\partial }{\partial y}   \left[ v(y) P^{surv}_{\tau}(y \vert x)  -D(y)  \frac{\partial P^{surv}_{\tau}(y\vert x)}{\partial y}     \right]
-  \lambda( y)   P^{surv}_{\tau}(y\vert x)
\label{jumpdiffabs}
\end{eqnarray}
with the initial condition at $\tau=0$
\begin{eqnarray}
 P^{surv}_{0}(y \vert x)  = \delta( y - x) 
\label{jumpdiffabsini}
\end{eqnarray}
The path-integral representation of the solution reads
\begin{eqnarray}
P^{surv}_{\tau}(  y  \vert x  )
 =   \int_{z(0)=x}^{z(\tau)=y} {\cal D} z(.) \ e^{ - \displaystyle 
  \int_0^{\tau} ds \left( \lambda(z(s)) + \frac{ \left[ \dot z(s) - v(z(s)) \right]^2 }{4D(z(s)) }
 - \frac{  [D'(z(s)) ]^2}{16D(z(s))} + \frac{  D''(z(s)) }{4} + \frac{  v'(z(s)) }{2}
\right)  }
\label{psurv}
\end{eqnarray}
So the excursion kernel of Eq. \ref{Wesc} will be explicit whenever one can solve
the time-dependent absorbing dynamics of Eqs \ref{jumpdiffabs} \ref{jumpdiffabsini}
or equivalently compute the path-integral of Eq. \ref{psurv}.
Let us now describe some simple examples.


\subsection{ Simplifications of the excursion kernel $W(\tau ,  y  \vert x  ) $ for jump-drift models without diffusion $D(y)=0$ }

For jump-drift models without diffusion  $D(y)=0$, the path-integral of Eq. \ref{psurv} reduces to the single deterministic trajectory
$z_x (0 \leq s \leq \tau) $ satisfying the equation of motion with the drift $v(.)$ 
\begin{eqnarray}
\frac{d z_x (s)}{ds}  = v(z_x(s))
\label{determi}
\end{eqnarray}
and starting at the position $x$ at the initial time $s=0$
\begin{eqnarray}
z_x(0) && =x
\label{determiini}
\end{eqnarray}
The solution 
\begin{eqnarray}
P^{surv}_{\tau}(  y  \vert x  )
 =  \delta\left( y - z_x(\tau) \right)  e^{ - \displaystyle \int_0^{\tau} ds  \lambda(z_x(s)) }
\label{psurvsansdiff}
\end{eqnarray}
can be rewritten using the separation of variables $ds=\frac{d z}{v(z)}$ of Eq. \ref{determi}
as 
\begin{eqnarray}
P^{surv}_{\tau}(  y  \vert x  )
 =  \frac{ \delta\left( \tau - \int_x^{y}   \frac{ dz }{v(z) } \right)  }{ \vert v(y) \vert } \ \ e^{ - \displaystyle \int_x^{y} dz  \frac{ \lambda(z) }{v(z) } }
\label{psurvsansdiffspace}
\end{eqnarray}
So the excursion kernel of Eq. \ref{Wesc} reduces to 
\begin{eqnarray}
W (\tau ,  y  \vert x  )
 =  W^{sp}(  y  \vert x  )
  \delta\left( \tau - \int_x^{y}   \frac{ dz }{v(z) } \right) 
\label{Wescsansdiff}
\end{eqnarray}
where 
\begin{eqnarray}
W^{sp}(  y  \vert x  )
 =   \frac{  \lambda( y) }{ \vert v(y) \vert  } \ \ e^{ - \displaystyle \int_x^{y} dz  \frac{ \lambda(z) }{v(z) } }
 \theta \left( \int_x^{y}   \frac{ dz }{v(z) } \geq 0 \right) 
\label{Wspatial}
\end{eqnarray}
represents the spatial probability that an excursion ends at position $y$ if it starts at position $x$,
the last Heaviside function ensuring that the corresponding duration is positive $\tau \geq 0$.
In most models of interest, the drift $v(z)$ is a continuous function of the position $z$,
so that the sign of the velocity $v(x)$ at the initial point will determine
the sign of the velocity during the whole excursion and
whether the position of the end-point $y$ is greater or smaller than $x$.
Then Eq. \ref{Wspatial} can be rewritten more explicitly as
\begin{eqnarray}
W^{sp}(  y  \vert x  ) 
&& =  \frac{  \lambda( y) }{ \vert v(y) \vert  } \ \ e^{ - \displaystyle \int_x^{y} dz  \frac{ \lambda(z) }{v(z) } }
 \theta \left( (y-x) v(x) \geq 0 \right) 
\nonumber \\
&&  =  \theta \left( v(x) >  0 \right) \theta \left( y >  x \right) W^{sp}_+(  y  \vert x  )
 +  \theta \left( v(x) <  0 \right) \theta \left( y <  x \right) W^{sp}_-(  y  \vert x  )
\label{2cases}
\end{eqnarray}
in terms of the two cases :

(i) when the drift is positive $v(x)>0$ at the initial point $x$, Eq. \ref{Wspatial} reads more explicitly
\begin{eqnarray}
W^{sp}_+(  y  \vert x  )
 =   \frac{  \lambda( y) }{ v(y) } \ \ e^{ - \displaystyle \int_x^{y} dz  \frac{ \lambda(z) }{v(z) } }
 = - \frac{d}{dy} \ e^{ - \displaystyle \int_x^{y} dz  \frac{ \lambda(z) }{v(z) } }
\ \ \ \  \ \ {\rm for } \ \ y \geq x
\label{WWspatialp}
\end{eqnarray}
 with the normalization
\begin{eqnarray}
\int_x^{+\infty} dy W^{sp}_+(  y  \vert x  )
 =   1
\label{WWspatialpnorm}
\end{eqnarray}

(ii) When the drift is negative $v(x)<0$ at the initial point $x$, Eq. \ref{Wspatial} reads more explicitly
\begin{eqnarray}
 \nonumber \\
 W^{sp}_-(  y  \vert x  )
 =    \frac{  \lambda( y) }{ \vert v(y) \vert } \ \ e^{  \displaystyle - \int_y^x dz  \frac{ \lambda(z) }{ \vert v(z) \vert } }
 = \frac{d}{dy} \ e^{ - \displaystyle \int_y^x dz  \frac{ \lambda(z) }{ \vert v(z) \vert } }
\ \ \ \  \ \ {\rm for } \ \ y \leq x
\label{WWspatialm}
\end{eqnarray}
 with the normalization
\begin{eqnarray}
\int_{-\infty}^x dy W^{sp}_-(  y  \vert x  )
 =  1
\label{WWspatialnorm}
\end{eqnarray}

Three examples of jump-drift models without diffusion $D(x)=0$ 
 will be described in sections \ref{sec_reset0}, \ref{sec_blowup}, \ref{sec_rain}.


\subsection{ Simplifications of the excursion kernel $W(\tau ,  y  \vert x  ) $ when the jump rate is uniform $\lambda(x) = \lambda$ }

When the jump rate is uniform $\lambda(x) = \lambda$, the path-integral of Eq. \ref{psurv} reduces to
\begin{eqnarray}
P^{surv}_{\tau}(  y  \vert x  )
 =  e^{\displaystyle - \tau \lambda}  P^{free}_{\tau}(  y  \vert x  )
 \label{psurvfree}
\end{eqnarray}
where 
\begin{eqnarray}
P^{free}_{\tau}(  y  \vert x  )
 =   \int_{z(0)=x}^{z(\tau)=y} {\cal D} z(.) \ e^{ - \displaystyle 
  \int_0^{\tau} ds \left(  \frac{ \left[ \dot z(s) - v(z(s)) \right]^2 }{4D(z(s)) }
 - \frac{  [D'(z(s)) ]^2}{16D(z(s))} + \frac{  D''(z(s)) }{4} + \frac{  v'(z(s)) }{2}
\right)  }
\label{pfree}
\end{eqnarray}
is the free propagator of the diffusion process
\begin{eqnarray}
\frac{ \partial P^{free}_{\tau}(y \vert x)   }{\partial \tau}  && = 
 -   \frac{\partial }{\partial y}   \left[ v(y) P^{free}_{\tau}(y\vert x )  -D(y)  \frac{\partial P^{free}_{\tau}(y\vert x)}{\partial y}     \right]
\nonumber \\
 P^{free}_{0}(y\vert x )  && = \delta( y - x) 
\label{freepropa}
\end{eqnarray}
normalized to unity at any time $\tau$
\begin{eqnarray}
\int dy P^{free}_{\tau}(  y  \vert x  )
 =  1
\label{pfreenorma}
\end{eqnarray}
So the kernel of Eq. \ref{Wesc} is factorized 
\begin{eqnarray}
W(\tau ,  y  \vert x  )
 = E^{exc} (\tau)  P^{free}_{\tau}(  y  \vert x  )
\label{Wescfactor}
\end{eqnarray}
into the normalized exponential probability to see the duration $\tau \in ]0,+\infty[$
\begin{eqnarray}
E^{exc} (\tau) = \lambda  e^{\displaystyle - \tau \lambda}  
\label{tauexpo}
\end{eqnarray}
and into the free propagator $P^{free}_{\tau}(  y  \vert x  )$ discussed above.
In conclusion, when the jump is uniform $\lambda(x) = \lambda$,
the excursion kernel $W(\tau ,  y  \vert x  ) $ is explicit 
whenever the free propagator $P^{free}_{\tau}(  y  \vert x  ) $ is known.
Two example of jump-diffusion models with uniform jump rate $\lambda(x) = \lambda $
will be described in sections \ref{sec_diffusionresetR} and \ref{sec_OUavecjumpHz}.


\section{ Large deviations for the empirical excursions between consecutive jumps }

\label{sec_largedevexcursion}

\subsection{ Density $ q(\tau, y  , x)$ of empirical excursions between two consecutive jumps  }

For the jump events, we have already introduced the empirical jump-flow $Q(x,y) $ in Eq. \ref{Qflow}
that can be rewritten more explicitely with the notations of Eqs \ref{yibefore} \ref{xiafter}
\begin{eqnarray}
Q(x,y)  = \frac{1}{T} \sum_{i=1}^N   \delta( x -  x(t_{i}^+) ) \delta( y -  x(t_{i}^-) ) 
\label{Qflowsumi}
\end{eqnarray}
with the corresponding in-flow and out-flow of Eq. \ref{Qflowpm}
\begin{eqnarray}
Q^+(x) \equiv  \int d y Q(x,y)  =   \frac{1}{T} \sum_{i=1}^N   \delta( x -  x(t_{i}^+) ) 
\nonumber \\
Q^-(y) \equiv  \int d x Q(x,y)  =   \frac{1}{T} \sum_{i=1}^N   \delta( y -  x(t_{i}^-) ) 
\label{Qflowinout}
\end{eqnarray}
In this section, we are interested into the empirical density
of the excursions between two consecutive jumps 
\begin{eqnarray}
q(\tau, y  , x)  \equiv \frac{1}{T} \sum_{i=1}^N  \delta(\tau-(t_{i+1}-t_i) ) \delta( y -  x(t_{i+1}^-) ) 
\delta( x -  x(t_{i}^+) )
\label{qexc}
\end{eqnarray}
that also contains the information on the in-flow and out-flow of Eq. \ref{Qflowinout} 
after integration over one position and over the duration $\tau$
\begin{eqnarray}
Q^+(x) && = \frac{1}{T} \sum_{i=1}^N   \delta( x -  x(t_{i}^+) ) 
= \int_{0}^{+\infty} d \tau \int d y \ q(\tau,y , x)
\nonumber \\
Q^-(y) && = \frac{1}{T} \sum_{i=0}^{N-1}   \delta( y -  x(t_{i+1}^-) ) 
= \int_{0}^{+\infty} d \tau \int d x \ q(\tau, y , x)
\label{Qflowinoutfromqexc}
\end{eqnarray}
The total density of excursions of duration $\tau$ can be obtained via the integration over the two positions
\begin{eqnarray}
q(\tau) \equiv \int d x \int d y q(\tau, y  ; x)  = \frac{1}{T} \sum_{i=1}^N  \delta(\tau-(t_{i+1}-t_i) )
 \label{qexctaualone}
\end{eqnarray}
The sum of the durations $\tau_i=t_{i+1}-t_i$ of all the excursions determines the normalization
\begin{eqnarray}
1 = \frac{1}{T} \sum_{i=1}^N  (t_{i+1}-t_i) =  \int_{0}^{+\infty} d \tau \tau q(\tau)   
= \int_{0}^{+\infty} d \tau \tau \int d x \int d y q( \tau , y , x)
 \label{totaltime}
\end{eqnarray}
while the total density $n=\frac{N}{T} $ of jumps of Eq. \ref{ndensity} corresponds to the total density of excursions
\begin{eqnarray}
n=\frac{N}{T}  = \int d x \int d y Q(x,y)  = \int d x \ Q^+(x) = \int d y \ Q^-(y) 
=  \int_{0}^{+\infty} d \tau \int d x \int d y \ q(\tau, y , x)
\label{ndensityesc}
\end{eqnarray}

It is important to stress that even the empirical current $j(.) $ 
can actually be reconstructed via the formula
\begin{eqnarray}
  j(z) =  \int_{0}^{+\infty} d \tau \left[ 
   \int_z^{+\infty}  d y \int_{-\infty}^z d x \ q(\tau, y , x)
-  \int_{-\infty}^z dy \int_z^{+\infty} dx \ q(\tau, y , x) 
  \right]
\label{jrecover}
\end{eqnarray}
whose derivative coincides with the stationarity condition of Eq \ref {statio}
\begin{eqnarray}
  j' (z) && =  \int_{0}^{+\infty} d \tau \left[ 
    - \int_{-\infty}^z d x \ q(\tau, z , x)
  +  \int_z^{+\infty}  d y  \ q(\tau, y , z)
-  \int_z^{+\infty} dx \ q(\tau, z , x)
 + \int_{-\infty}^z dy  \ q(\tau, y , z) 
  \right]
  \nonumber \\
  && = \int_{0}^{+\infty} d \tau \left[ 
  \int_{-\infty}^{+\infty}  d y  \ q(\tau, y , z)
 -  \int_{-\infty}^{+\infty} dx \ q(\tau, z , x)
  \right] = Q^+(z)-Q^-(z)
\label{jrecoverderi}
\end{eqnarray}
The physical meaning of Eq. \ref{jrecover} is that
the forward excursions $q(\tau, y , x) $ with $y>x$ correspond to a positive current contribution at 
any point of the interval $z \in ]x,y[$ (first term),
while the backward excursions $q(\tau, y , x) $ with $y<x$ correspond to a negative current contribution at 
any point of the interval $z \in ]y,x[$ (second term).

Finally, for the empirical density $\rho(x)$, one must distinguish two cases :

(i) for jump-drift models without diffusion $D(x) = 0 $, 
the empirical density can be reconstructed via $\rho(x)=\frac{j(x)}{v(x)}$ (Eq. \ref{jrhonodiff}) from the empirical current $j(x)$ of Eq. \ref{jrecover} as will be discussed in more detail in the subsection \ref{subsec_excnodiff}.

(ii) for jump-diffusion models with non-vanishing diffusion $D(x) \ne 0$, 
the empirical density $\rho(.)$ cannot be reconstructed from the empirical excursions $q(\tau,y,x)$ alone,
i.e. some information on the position during the excursions has been lost.


\subsection{ Large deviations for the empirical jumps and for the empirical excursions between jumps }

The joint distribution of the density $n$ of jumps,
of the in-flow $Q^+(.)  $, of the out-flow $Q^-(.)  $,
of jump-flow $Q(., .) $, 
and of the density $q(.,.,.)$ of excursions between jumps, with its partial density $q(\tau)$ of the duration $\tau$
follows the large deviation form
\begin{eqnarray}
&& {\cal P}_T[ n; Q^{\pm}(.) ;  Q(.,.) ; q(.) ; q(.,.,.)   ] \opsimeq_{T \to +\infty}  
{\cal C}[ n; Q^{\pm}(.); Q(.,.) ; q(.) ; q(.,.,.)  ] 
 e^{\displaystyle - T  {\cal I} [  Q^{\pm}(.); Q(.,.) ; q(.,.,.)  ]    }
\label{ld2.75}
\end{eqnarray}
The constraints
\begin{eqnarray}
 && {\cal C}[ n; Q^{\pm}(.); Q(.,.) ; q(.) ; q(.,.,.)   ]  =
 \delta \left(  \int d y \ Q^-(y) - n     \right)    \delta \left(  \int d x \ Q^+(x) - n     \right)   
       \nonumber \\ &&
  \delta \left( \int_{0}^{+\infty} d \tau \tau q(\tau)  -1  \right)
   \left[  \prod_{\tau>0} \delta \left(  \int d y \int dx \ q(\tau,y , x) - q(\tau)     \right)  \right]              
       \nonumber \\ &&      
 \left[ \prod_{ x } \delta \left(  \int d y \  Q(x,y) - Q^+(x)     \right) 
 \delta \left( \int_{0}^{+\infty} d \tau \int d y \ q(\tau,y , x) - Q^+(x)     \right)
  \right]
\nonumber \\
&&\left[ \prod_{ y } \delta \left(  \int d x \ Q(x,y) - Q^-(y) \right) 
  \delta \left(  \int_{0}^{+\infty} d \tau \int d x \ q(\tau, y , x)  - Q^-(y)  \right)  \right]
\label{c2.75}
\end{eqnarray}
can be understood as follows :
 the first line contains the definition of the total density $n$ of Eq. \ref{ndensityesc},
 the second line contains the normalization of Eq. \ref{totaltime}
 with the definition of $q(\tau)$ of Eq. \ref{qexctaualone},
while the two last lines contain the definitions of 
the in-flow $Q^+(.)$ and of the out-flow $Q^-(.)$ of Eqs \ref{Qflowinout} and \ref{Qflowinoutfromqexc}.
The rate function corresponds to the alternate Markov chain of Eq. \ref{alternatemarkov}
and contains the two corresponding contributions :
\begin{eqnarray}
  {\cal I} [ Q^{\pm}(.); Q(.,.) ; q(.,.,.)  ]   =  {\cal I}^{[\Pi]} [  Q^{-}(.); Q(.,.)   ] +  {\cal I}^{[W]} [  Q^{+}(.) ; q(.,.,.)  ] 
\label{rate2.75}
\end{eqnarray}

(i) The first contribution involving the jump kernel $\Pi( x \vert y)$ 
\begin{eqnarray}
 {\cal I}^{[\Pi]} [  Q^{-}(.); Q(.,.)   ] =  \int d x \int d y
Q(x,y) \ln \left( \frac{ Q(x,y)  }{   \Pi( x \vert y)  Q^-(y) }  \right) \equiv  I^{[\Pi]}_{2.5} [  Q^-(.), Q(.,.) ]
\label{rate2.75jump}
\end{eqnarray}
coincides with the contribution $  I^{[\Pi]}_{2.5} [  Q^-(.), Q(.,.) ] $ of Eq. \ref{rate2.5jumppi}
discussed previously.

(ii) The second contribution involving the excursion kernel $W(\tau,  y  \vert x  )$ of Eq. \ref{Wesc}
\begin{eqnarray}
 {\cal I}^{[W]} [  Q^{+}(.) ; q(.,.,.)  ] 
= \int_0^{+\infty} d \tau \int d x \int d y \ 
 q(\tau,y , x)  \ln \left( \frac{ q(\tau,y , x) }{ W(\tau,  y  \vert x  ) Q^+(x)  }   \right)
\label{rate2.75exc}
\end{eqnarray}
takes into account the diffusion coefficient $D(.)$, the drift $v(.)$ and the jump rate $\lambda(.)$
that determine the excursion kernel $W(\tau,  y  \vert x  ) $.


\subsection{ Steady state properties from the point of view of the jump events only }

From the point of view of the jump events only, the steady state properties make the rate functions of Eq. \ref{rate2.75jump}
and Eq. \ref{rate2.75exc} vanish
\begin{eqnarray}
Q_*(x,y)  && =   \Pi( x \vert y)  Q^-_*(y) 
\nonumber \\
q_*(\tau,y , x) && = W(\tau,  y  \vert x  ) Q^+_*(x)  
\label{ratevanish}
\end{eqnarray}
and satisfy all the constraints of Eq. \ref{c2.75}
\begin{eqnarray}
1 &&  = \int_{0}^{+\infty} d \tau \tau \int d x \int d y q_*(\tau,y , x)  
 = \int_{0}^{+\infty} d \tau \tau \int d x \int d y W(\tau,  y  \vert x  ) Q^+_*(x) 
\nonumber \\
n_* && =   \int d y \ Q^-_*(y)
\nonumber \\
n_* && = \int d x \ Q^+_*(x) 
       \nonumber \\ 
Q^+_*(x)      && =  \int d y \  Q_*(x,y) =  \int d y   \Pi( x \vert y)  Q^-_*(y) 
\nonumber \\
Q^-_*(y)  && =  \int_{0}^{+\infty} d \tau \int d x \ q_*(\tau,y , x) 
= \int_{0}^{+\infty} d \tau \int d x W(\tau,  y  \vert x  ) Q^+_*(x) 
\label{c2.75st}
\end{eqnarray}
while the two remaining constraints are automatically satisfied as a consequence of 
the normalizations of Eqs \ref{normaPi} and \ref{normaWesc}
\begin{eqnarray}
Q^+_*(x)      && =\int_{0}^{+\infty} d \tau \int d y \ q^*_{\tau} (y , x) 
= \left[ \int_{0}^{+\infty} d \tau \int d y W(\tau,  y  \vert x  ) \right] Q^+_*(x) = Q^+_*(x)
\nonumber \\
Q^-_*(y)  && =  \int d x \ Q_*(x,y)  =  \left[ \int d x   \Pi( x \vert y) \right]  Q^-_*(y) = Q^-_*(y)
\label{auto}
\end{eqnarray}

In summary, the steady state properties of the jump events alone can be found as follows.
The in-flow $ Q^+_*(x) $ is the steady state of the global composite kernel
\begin{eqnarray}
Q^+_*(x)   =  \int d x'
\left[   \int d y   \Pi( x \vert y)   \int_{0}^{+\infty} d \tau  W(\tau,  y  \vert x'  ) \right] Q^+_*(x') 
\label{q+stalone}
\end{eqnarray}
with the normalization
\begin{eqnarray}
1  = \int_{0}^{+\infty} d \tau \tau \int d x \int d y W(\tau,  y  \vert x  ) Q^+_*(x) 
\label{normaq+stalone}
\end{eqnarray}
The jump density $n_*$ and the out-flow $Q^-*(y)$ can be then computed via
\begin{eqnarray}
n_* && = \int d x \ Q^+_*(x) 
\nonumber \\
Q^-_*(y)  && = \int_{0}^{+\infty} d \tau \int d x W(\tau,  y  \vert x  ) Q^+_*(x) 
\label{nqmfromq+}
\end{eqnarray}


\subsection{ Simplifications of empirical excursions for  jump-drift models without diffusion $D(y)=0$ }

\label{subsec_excnodiff}

For  jump-drift models without diffusion $D(y)=0$, the factorization of Eq. \ref{Wescsansdiff}
for the excursion kernel
\begin{eqnarray}
W(\tau,  y  \vert x  ) =  W^{sp}(  y  \vert x  )  \delta\left( \tau - \int_x^{y}   \frac{ dz }{v(z) } \right) 
\label{Wescsansdifffactor}
\end{eqnarray}
yields that the empirical excursions display the same factorization
between the spatial part $q^{sp} (y , x) $ and the delta function for the corresponding duration $\tau$
\begin{eqnarray}
q(\tau,y , x) = q^{sp} (y , x) \delta\left( \tau - \int_x^{y}   \frac{ dz }{v(z) } \right) 
\label{qexcnodiff}
\end{eqnarray}
The spatial part $q^{sp} (y , x) $ contains the same Heaviside function $\theta \left( \int_x^{y}   \frac{ dz }{v(z) } > 0 \right) $
as Eq. \ref{Wspatial}
\begin{eqnarray}
q^{sp} (y , x) = \int_0^{+\infty} d \tau q(\tau,y , x) = q^{sp} (y , x)  \theta \left( \int_x^{y}   \frac{ dz }{v(z) } > 0 \right) 
\label{qspatial}
\end{eqnarray}
As a consequence, one can rewrite in terms of the spatial 
part $q^{sp} (y , x) $
both 
the rate function of Eq. \ref{rate2.75exc}
\begin{eqnarray}
&& {\cal I}^{[W^{sp}]}_{[D(.)=0]} [  Q^{+}(.) ; q^{sp}(.,.)  ] 
 =  \int d x \int d y \ \theta \left( \int_x^{y}   \frac{ dz }{v(z) } \geq 0  \right)
 q^{sp} (y , x)  \ln \left( \frac{ q^{sp} (y , x) }{ W^{sp}(  y  \vert x  ) Q^+(x)  }   \right)
\label{rate2.75excnodiff}
\end{eqnarray}
and the constraints of Eq. \ref{c2.75} 
\begin{eqnarray}
 && {\cal C}_{[D(.)=0]}[ n; Q^{\pm}(.); Q(.,.) ; q^{sp}(.,.)   ]  =
 \delta \left(  \int d y \ Q^-(y) - n     \right)    \delta \left(  \int d x \ Q^+(x) - n     \right)   
       \nonumber \\ &&
\delta \left( \int d x \int d y q^{sp} (y , x) \left[ \int_x^{y}   \frac{ dz }{v(z) }  \right] \theta \left( \int_x^{y}   \frac{ dz }{v(z) } \geq 0  \right) -1  \right)
       \nonumber \\ &&      
 \left[ \prod_{ x } \delta \left(  \int d y \  Q(x,y) - Q^+(x)     \right) 
 \delta \left(  \int d y \ q^{sp}(y , x) \theta \left( \int_x^{y}   \frac{ dz }{v(z) } \geq 0  \right)- Q^+(x)     \right)
  \right]
\nonumber \\
&&\left[ \prod_{ y } \delta \left(  \int d x \ Q(x,y) - Q^-(y) \right) 
  \delta \left(   \int d x \ q^{sp}( y , x) \theta \left( \int_x^{y}   \frac{ dz }{v(z) } \geq 0  \right) - Q^-(y)  \right)  \right]
\label{c2.75nodiffsp}
\end{eqnarray}
The reconstruction of the empirical current via Eq. \ref{jrecover} becomes
\begin{eqnarray}
  j(z) = \theta( v(z)>0)   \int_z^{+\infty}  d y \int_{-\infty}^z d x \ q^{sp} ( y , x) 
 - \theta( v(z)<0) \int_{-\infty}^z dy \int_z^{+\infty} dx \ q^{sp} ( y , x) 
\label{jrecoversp}
\end{eqnarray}
while the empirical density can be then obtained from Eq. \ref{jrhonodiff}
in the absence of diffusion $D(x)=0$
\begin{eqnarray}
\rho(z)  && = \frac{ j(z) }{v(z) }   = 
 \frac{ 1 }{v(z) } \left[ 
 \theta( v(z)>0)   \int_z^{+\infty}  d y \int_{-\infty}^z d x \ q^{sp} ( y , x) 
 - \theta( v(z)<0) \int_{-\infty}^z dy \int_z^{+\infty} dx \ q^{sp} ( y , x) 
  \right]
  \nonumber \\
  && = \frac{1}{\vert v(z) \vert }
  \left[ 
 \theta( v(z)>0)   \int_z^{+\infty}  d y \int_{-\infty}^z d x \ q^{sp} ( y , x) 
 + \theta( v(z)<0) \int_{-\infty}^z dy \int_z^{+\infty} dx \ q^{sp} ( y , x) 
  \right]
\label{rhorecovernodiff}
\end{eqnarray} 
Note that the normalization of this expression for the empirical density 
coincides with the second line of Eq. \ref{c2.75nodiffsp} concerning the durations of excursions
\begin{eqnarray}
1 && = \int_{-\infty}^{+\infty} dz \rho(z)     = 
  \int_{-\infty}^{+\infty}  d y \int_{-\infty}^{y} d x \ q^{sp} ( y , x) \int_x^y  \frac{ dz }{v(z) } 
- \int_{-\infty}^{+\infty} dy \int_{y}^{+\infty} dx \ q^{sp} ( y , x) \int_y^x \frac{ dz }{v(z) } 
\nonumber \\
&& =\int d x \int d y \ q^{sp} (y , x) \left[ \int_x^{y}   \frac{ dz }{v(z) }  \right] \theta \left( \int_x^{y}   \frac{ dz }{v(z) } \geq 0  \right)
\label{normarhorecovernodiff}
\end{eqnarray} 

Putting everything together, one obtains that for jump-drift models without diffusion $D(x)=0$,
the joint distribution of the empirical density $\rho(.)$, the jump density $n$, 
the in-flow and the out-flow $Q^{\pm}(.)$, the jump flow $Q(.,.)$, and the spatial excursion density $q^{sp}(.,.)$
follow the large deviations at Level 2.75
\begin{eqnarray}
 P_T^{2.75[D(.)=0]}[ \rho(.), n,Q^{\pm}(.),Q(.,.),q^{sp} ( . , .)]  
   \opsimeq_{T \to +\infty}  
{\cal C}_{2.75}[ \rho(.), n,Q^{\pm}(.),Q(.,.),,q^{sp} ( . , .)] 
e^{ \displaystyle -T  {\cal I}_{2.75} [ Q^{\pm}(.),Q(.,.),q^{sp}(.,.)] }
  \nonumber \\ 
\label{ld2.75nodiff}
\end{eqnarray}
The constraints ${\cal C}_{2.75}[ \rho(.), n,Q^{\pm}(.),Q(.,.),,q^{sp} ( . , .)]  $ at Level 2.75 include 
 the constraints $ C^{[D(.)=0]}_{2.5}[ \rho(.), n,Q^{\pm}(.),Q(.,.)] $ of the Level 2.5 of Eq. \ref{c2.5nodiff}
and contains in addition the definitions of the in-flow $Q^+$ and of the out-flow $Q^-$ in terms of 
the spatial excursions $q^{sp}(.,.)$
\begin{eqnarray}
&& {\cal C}_{2.75}[ \rho(.), n,Q^{\pm}(.),Q(.,.),,q^{sp} ( . , .)]  
= C^{[D(.)=0]}_{2.5}[ \rho(.), n,Q^{\pm}(.),Q(.,.)]  
 \label{c2.75nodiff}  
      \\ &&      
 \left[ \prod_{ x }    \delta \left(  \int d y \ \theta \left( \int_x^{y}   \frac{ dz }{v(z) } \geq 0  \right)   q^{sp}(y , x) - Q^+(x)     \right)  \right]
\left[ \prod_{ y }   \delta \left(   \int d x \ \theta \left( \int_x^{y}   \frac{ dz }{v(z) } \geq 0  \right)   q^{sp}( y , x)  - Q^-(y)  \right)  \right]
 \end{eqnarray}
The rate function at Level 2.75 contains the jump contribution of Eq. \ref{rate2.75jump}
and the excursion contribution of Eq. \ref{rate2.75excnodiff}
\begin{eqnarray}
 {\cal I}_{2.75} [ Q^{\pm}(.),Q(.,.),q^{sp}(.,.)]={\cal I}^{[\Pi]} [  Q^{-}(.); Q(.,.)   ] +  {\cal I}^{[W]}_{[D(.)=0]} [  Q^{+}(.) ; q^{sp}(.,.)  ]
\label{rate2.75sp}
\end{eqnarray}

The comparison with the Level 2.5 of Eq. \ref{ld2.5nodiff} yields that the
conditional probability to see spatial excursions
$q^{sp} ( . , .) $ once all the other empirical observables are given reduces to
\begin{eqnarray}
&& {\cal P}_T^{conditional} [ q^{sp} ( . , .) \vert \rho(.), n,Q^{\pm}(.),Q(.,.) ]  \equiv \frac{ {\cal P}_T^{2.75}[ \rho(.), j(.),n,Q^{\pm}(.),Q(.,.),q^{sp} ( . , .)]  }
{ P_T^{2.5}[ \rho(.), n,Q^{\pm}(.),Q(.,.)]  } 
\nonumber \\
&&  \opsimeq_{T \to +\infty}  
 \left[ \prod_{ x }    \delta \left(  \int d y \ q^{sp}(y , x)\theta \left( \int_x^{y}   \frac{ dz }{v(z) } \geq 0  \right)   - Q^+(x)     \right)  \right]
\left[ \prod_{ y }   \delta \left(   \int d x \ q^{sp}( y , x)\theta \left( \int_x^{y}   \frac{ dz }{v(z) } \geq 0  \right)    - Q^-(y)  \right)  \right]
  \nonumber \\
&&
e^{ \displaystyle -T   {\cal I}^{conditional} [ \rho(.), Q^{\pm}(.),q^{sp}(.,.)]}
\label{ld2.75condi}
\end{eqnarray}
where the conditional rate function reads
\begin{eqnarray}
&&  {\cal I}^{conditional} [ \rho(.), Q^{\pm}(.),q^{sp}(.,.)] 
   =
  {\cal I}_{2.75} [ Q^{\pm}(.),Q(.,.),q^{sp}(.,.)] - I_{2.5}^{[D(.)=0]}[ \rho(.), Q^{\pm}(.),Q(.,.)  
 \label{rate2.75conditional}
 \\
  && =  {\cal I}^{[W^{sp}]}_{[D(.)=0]} [  Q^{+}(.) ; q^{sp}(.,.)  ]
  - I^{[\lambda]}_{2.5} [ \rho(.), Q^-(.)]  
  \nonumber \\
  && = 
   \int d x \int d y \ \theta \left( \int_x^{y}   \frac{ dz }{v(z) } \geq 0  \right)
 q^{sp} (y , x)  \ln \left( \frac{ q^{sp} (y , x) }{ W^{sp}(  y  \vert x  ) Q^+(x)  }   \right)
   - \int d y \left[ Q^-(y) \ln \left( \frac{  Q^-(y)  }{    \lambda(y)   \rho(y) }  \right) 
 -  Q^-(y) +   \lambda(y)   \rho(y)  \right]
  \nonumber
\end{eqnarray}

The explicit form of the spatial kernel $W^{sp}(  y  \vert x  ) $ of Eq. \ref{Wspatial},
can be translated for
 the effective spatial kernel $ \hat W^{sp}(  y  \vert x  ) $ associated to the effective jump rate 
$\hat \lambda(y) \equiv  \frac{  Q^-(y)  }{   \rho(y) }  $ of Eq. \ref{hatlambda}
\begin{eqnarray}
{\hat W}^{sp}(  y  \vert x  )
 =   \frac{  \hat \lambda( y) }{ \vert v(y) \vert  } \ \ e^{ - \displaystyle \int_x^{y} dz  \frac{ \hat \lambda(z) }{v(z) } }
 \theta \left( \int_x^{y}   \frac{ dz }{v(z) } \geq 0 \right) 
 =  \frac{  Q^-( y) }{ \rho(y) \vert v(y) \vert  } \ \ e^{ - \displaystyle \int_x^{y} dz  \frac{ Q^-(z) }{ \rho(z) v(z) } }
 \theta \left( \int_x^{y}   \frac{ dz }{v(z) } \geq 0 \right) 
\label{hatWspatial}
\end{eqnarray}
This effective spatial kernel $ \hat W^{sp}(  y  \vert x  ) $ is useful to rewrite the 
conditional rate function of Eq. \ref{rate2.75conditional} as
\begin{eqnarray}
&&  {\cal I}^{conditional} [ \rho(.), Q^{\pm}(.),q^{sp}(.,.)] 
 = {\cal I}^{[\hat W]}_{[D(.)=0]} [  Q^{+}(.) ; q^{sp}(.,.)  ]
  \nonumber \\
  && = 
   \int d x \int d y \ \theta \left( \int_x^{y}   \frac{ dz }{v(z) } \geq 0  \right)
 q^{sp} (y , x)  \ln \left( \frac{ q^{sp} (y , x) }{ \hat W^{sp}(  y  \vert x  ) Q^+(x)  }   \right)
   \nonumber \\
  && = 
   \int d x \int d y \ \theta \left( \int_x^{y}   \frac{ dz }{v(z) } \geq 0  \right)
 q^{sp} (y , x)  \ln \left( \frac{ q^{sp} (y , x) }{ 
 \frac{  Q^-( y) }{ \rho(y) \vert v(y) \vert  } \ \ e^{ -  \int_x^{y} dz  \frac{ Q^-(z) }{ \rho(z) v(z) } }
  Q^+(x)  }   \right)
  \label{rate2.75conditionalhat}
\end{eqnarray}
This factorized form shows that this conditional rate function vanishes for the optimal value
\begin{eqnarray}
q^{sp}_{opt} (y , x)= {\hat W}^{sp}(  y  \vert x  ) Q^+(x) 
 = Q^+(x)  \frac{  Q^-( y) }{ \rho(y) \vert v(y) \vert  } \ \ e^{ - \displaystyle \int_x^{y} dz  \frac{ Q^-(z) }{ \rho(z) v(z) } }
 \theta \left( \int_x^{y}   \frac{ dz }{v(z) } \geq 0 \right) 
\label{qspoptimal}
\end{eqnarray}
once all the other one-position empirical observables $[ \rho(.), Q^{\pm}(.)] $ are given.
Three examples of jump-drift models without diffusion $D(x)=0$ 
 will be described in sections \ref{sec_reset0}, \ref{sec_blowup}, \ref{sec_rain}.


\section{ Example: jump-drift process $[v(x)>0, \lambda(x) ]$ with origin resetting $\Pi( x \vert y)  =  \delta(x)$    }

\label{sec_reset0}

In this section, we consider the positive jump-drift process $x(t) \geq 0$ without diffusion $D(x)=0$,
with the space-dependent positive velocity $v(x)>0$,
and the space-dependent jump rate $\lambda(x)$,
while the jump probability 
\begin{eqnarray}
\Pi( x \vert y)  = \delta \left(x \right)   
\label{deterreset}
\end{eqnarray}
describes the stochastic resetting towards the origin $x=0$ (see the review \cite{review_search} 
on stochastic resetting and references therein).
The large deviations at Level 2.5 for resetting towards the origin 
have been already discussed in detail in \cite{c_reset}
for discrete-time Markov chains, continuous-time Markov jump processes and diffusion processes.
However, the present example corresponding to the continuous-time continuous-space version of the Sisyphus Random Walk \cite{sisyphus} in an arbitrary space-dependent landscape parametrized by drift $v(x)>0$, and the jump rate $\lambda(x)$,
is useful here as the simplest possible application of the present general formalism,
and as a comparison for the more complicated examples considered in the next sections.


\subsection{ Normalizability of the steady state $\rho_*(x)$}

The steady-state $\rho_*(x)$ satisfying Eq. \ref{jumpdiffst}
\begin{eqnarray}
  \frac{ d }{ d x}   \left[ v(x) \rho_*(x )   \right]
+  \lambda( x)   \rho_*(x) =\delta ( x )  \int_0^{+\infty} d y    \lambda( y)  \rho_*(y) 
\label{eqstresetorigin}
\end{eqnarray}
reads
\begin{eqnarray}
 \rho_*(x )  =  \rho_*(0 ) \frac{v(0)}{v(x)} e^{- \displaystyle\int_0^x dy \frac{\lambda (y)} {v(y) } }
\label{stresetorigin}
\end{eqnarray}
It is normalizable if the following integral involving the drift $v(.)$ and the jump rate $\lambda(.)$ converges 
\begin{eqnarray}
1=\int_0^{+\infty} dx  \rho_*(x )  =
  \rho_*(0 )v(0)
  \int_0^{+\infty} \frac{dx}{v(x)} e^{- \displaystyle\int_0^x dy \frac{\lambda (y)} {v(y) } }
\label{normastresetorigin}
\end{eqnarray}


\subsection{ Large deviations at Level 2.5  }

For the present model, the large deviations at Level 2.5 are  greatly simplified
because there is no diffusion $D(x)=0$ (Eq. \ref{ld2.5nodiff}) and because 
the jump probability is deterministic $ \Pi( x \vert y)  = \delta \left(x \right)   $ (Eq. \ref{ld2.5diffpideter}).
As a consequence, one obtains that 
the joint distribution of the empirical density $\rho(.)$
and of the empirical out-flow $Q^-(.)$ with the corresponding density $n$ reads
\begin{eqnarray}
 P_T[ \rho(.), n,Q^{-}(.)] &&  \opsimeq_{T \to +\infty}  
 \delta \left(\int_0^{+\infty} d x \rho(x) -1  \right)  \delta \left(  \int_0^{+\infty} d y \ Q^-(y) - n     \right)   
\left[  \prod_{x\geq 0 } \delta \left(  \frac{d}{dx} \left[  \rho(x) v(x)\right] + Q^-(x)   -  n \delta(x)  \right) \right]    
\nonumber \\
&& 
e^{- \displaystyle T \int_0^{+\infty} d y \left[ Q^-(y) \ln \left( \frac{  Q^-(y)  }{    \lambda(y)   \rho(y) }  \right) 
 -  Q^-(y) +   \lambda(y)   \rho(y)  \right]}
\label{ld2.5resetorigin}
\end{eqnarray}
while the in-flow $Q^+(.)$ can be obtained from the density $n$ alone
\begin{eqnarray}
Q^+(x)= n \delta(x) 
  \label{Qpdelta}
\end{eqnarray}
and the jump-flow $Q(.,.)$ can be computed from the out-flow $Q^-(.)$ alone
\begin{eqnarray}
Q(x,y)=  \delta(x) Q^-(y)
  \label{Qdeltaqm}
\end{eqnarray}


\subsubsection{ Large deviations at Level 2 for the empirical density $\rho(.)$ alone  }

One can use the stationarity constraint in Eq. \ref{ld2.5resetorigin}
to eliminate the out-flow $Q^-(.)$
in terms of  the empirical density $\rho(.)$ for $x>0$
\begin{eqnarray}
Q^-(x)= - \frac{d}{dx} \left[  \rho(x) v(x)\right]   
  \label{rhoqreset0elimiQ}
\end{eqnarray}
The jump density $n$ is then related to the empirical density $\rho(x=0)$ at the origin
\begin{eqnarray}
n = \int_0^{+\infty} dx Q^-(x)=  \rho(0) v(0) 
  \label{rhoqreset0elimiQn}
\end{eqnarray}
So one obtains that the large deviations at Level 2 for the empirical density $\rho(.)$ alone reads
\begin{eqnarray}
 P_T[ \rho(.)] &&  \opsimeq_{T \to +\infty}  
 \delta \left(\int_0^{+\infty} d x \rho(x) -1  \right)    e^{- \displaystyle T I_2[ \rho(.)] }
\label{ld2resetorigin}
\end{eqnarray}
where the rate function at Level 2 reads
\begin{eqnarray}
&& I_2[ \rho(.)]
\equiv \int_0^{+\infty} d x \left[ \left( - \frac{d}{dx} \left[  \rho(x) v(x)\right]   \right) \ln \left( \frac{ \left( - \frac{d}{dx} \left[  \rho(x) v(x)\right]   \right)   }{    \lambda(x)   \rho(x) }  \right) 
+ \frac{d}{dx} \left[  \rho(x) v(x)\right]   +   \lambda(x)   \rho(x)  \right]
\nonumber \\
&& 
= \int_0^{+\infty} d x  \left( - \frac{d}{dx} \left[  \rho(x) v(x)\right]   \right) \ln \left( \frac{ \left( - \frac{d}{dx} \left[  \rho(x) v(x)\right]   \right)   }{    \frac{\lambda(x)}{v(x) }    }  \right) 
+ \int_0^{+\infty} d x  \frac{d}{dx}\left(  \left[  \rho(x) v(x)\right]   \ln \left[    \rho(x) v(x)   \right] \right)
  + \int_0^{+\infty} d x  \lambda(x)   \rho(x)  
  \nonumber \\
 && 
= \int_0^{+\infty} d x  \left( - \frac{d}{dx} \left[  \rho(x) v(x)\right]   \right) \ln \left( \frac{ \left( - \frac{d}{dx} \left[  \rho(x) v(x)\right]   \right)   }{    \frac{\lambda(x)}{v(x) }    }  \right) 
-  \left[  \rho(0) v(0)\right]   \ln \left[    \rho(0) v(0)   \right] 
  + \int_0^{+\infty} d x  \lambda(x)   \rho(x)   
\label{rate2resetorigin}
\end{eqnarray}


\subsubsection{ Large deviations for the large deviations for the out-flow $Q^-(.)$ and the density $n$ alone  }

One can instead use the stationarity constraint to eliminate the empirical density $\rho(.)$
in terms of the out-flow $Q^-(.)$
\begin{eqnarray}
  \rho(x)  = \frac{1}{v(x)} \int_x^{+\infty} dy Q^-( y) 
  \label{rhoqreset0}
\end{eqnarray}
The normalization of the empirical density becomes
\begin{eqnarray}
\int_0^{+\infty} d x  \rho(x)  =\int_0^{+\infty} dy Q^-( y) \int_{0}^y \frac{dx}{v(x)} 
  \label{normarhoqreset0}
\end{eqnarray}
So one obtains the large deviation form
\begin{eqnarray}
 P_T[  n,Q^{-}(.)] &&  \opsimeq_{T \to +\infty}  
 \delta \left(\int_0^{+\infty} dy Q^-( y) \int_{0}^y \frac{dx}{v(x)} 
   -1  \right)  \delta \left(  \int_0^{+\infty} d y \ Q^-(y) - n     \right)    
e^{- \displaystyle T I[  n,Q^{-}(.)]}
\label{ldqout}
\end{eqnarray}
with the rate function translated from Eq. \ref{rate2resetorigin}
\begin{eqnarray}
 I[ n,Q^{-}(.)] 
&& = \int_0^{+\infty} d x  Q^-(x) \ln \left( \frac{ Q^-(x)   }{    \frac{\lambda(x)}{v(x) }    }  \right) 
-  n \ln n 
  + \int_0^{+\infty} d x     \frac{\lambda(x)}{v(x)} \int_x^{+\infty} dy Q^-( y) 
  \nonumber \\
&&  =
 \int_0^{+\infty} d y  Q^-(y) \ln \left( \frac{ Q^-(y)   }{    \frac{\lambda(y)}{v(y) }    }  \right) 
-  n \ln n 
  +  \int_0^{+\infty} dy Q^-( y) \int_0^{y} d x     \frac{\lambda(x)}{v(x)}
    \nonumber \\
&&  =
 \int_0^{+\infty} d y  Q^-(y) \ln \left( \frac{ Q^-(y)   }{    \frac{\lambda(y)}{v(y) }  e^{- \int_0^{y} d x     \frac{\lambda(x)}{v(x)}}  n }   \right) 
\label{rate2resetoriginQ}
\end{eqnarray}


\subsection{ Excursions between jumps }

The excursion kernel reduces to Eq. \ref{Wescsansdiff} for $x=0$ and $y \geq 0$
\begin{eqnarray}
W (\tau ,  y  \vert 0  )
 =  W^{sp}_+(  y  \vert 0  )  \delta\left( \tau - \int_0^{y}   \frac{ dz }{v(z) } \right) 
\label{Wescsansdiffreset0}
\end{eqnarray}
with Eq. \ref{WWspatialp}
\begin{eqnarray}
W_+^{sp}(  y  \vert 0  )
 =   \frac{  \lambda( y) }{ v(y) } \ \ e^{ - \displaystyle \int_0^{y} dz  \frac{ \lambda(z) }{v(z) } }
 = - \frac{d}{dy} \ e^{ - \displaystyle \int_0^{y} dz  \frac{ \lambda(z) }{v(z) } }
\label{WWspatialpreset}
\end{eqnarray}
For the present model, the empirical excursions can be rewritten in terms of the out-flow $Q^-(.)$
\begin{eqnarray}
q^{sp}(  y  , x  ) && = Q^-(y) \delta(x)
\nonumber \\
q (\tau ,  y  , x  ) && =  Q^-(y) \delta(x)  \delta\left( \tau - \int_0^{y}   \frac{ dz }{v(z) } \right) 
\label{qexcreset0}
\end{eqnarray}
and thus do not contain additional information with respect to the previous subsections.


\section{ Example: jump-drift process $[v(x)= v, \lambda(x)= \lambda ]$ 
with $\Pi^{deter}( x \vert y)  =  \delta(x-\gamma y)$  }
 
 \label{sec_blowup}
 
 As an example where the new position $x$ after the jump follows some 
 non-trivial deterministic rule $x = \Phi( y)$ (Eq. \ref{jumpdeter}),
let us consider the positive jump-drift process $x(t) \geq 0$ without diffusion $D(x)=0$,
with the uniform positive velocity $v(x)=v>0$
and with the uniform jump rate $\lambda(y)=\lambda $,
while the jump probability of parameter $\gamma \in ]0,1[$ (for instance $\gamma=\frac{1}{2}$)
\begin{eqnarray}
\Pi^{deter}( x \vert y)  = \delta \left(x - \gamma y \right)   
\label{deterblow}
\end{eqnarray}
describes backward jumps from $y$ to $x=\gamma y <y$ \cite{blowup}.

 
 \subsection{ Steady state $\rho_*(x)$ via its moments }

The dynamics of Eq. \ref{jumpdiff}
\begin{eqnarray}
\frac{ \partial \rho_t(x)   }{\partial t} && = -   \frac{d}{dx}  \left[  v \rho_t( x )   \right]
-  \lambda    \rho_t( x) +  \lambda \int d y  \  \delta \left(x - \gamma y \right)   \rho_t(y)
\label{jumpdeterst1d}
\end{eqnarray}
yields that the integer moments
\begin{eqnarray}
\langle x^k \rangle_t \equiv \int_0^{+\infty} dx x^k \rho_t(x)
\label{defmoments}
\end{eqnarray}
satisfy the closed dynamical equations
\begin{eqnarray}
\frac{ \partial \langle x^k \rangle_t   }{\partial t} = k v   \langle x^{k-1} \rangle_t 
-  \lambda (1-\gamma^k) \langle x^k \rangle_t 
\label{dynmoments}
\end{eqnarray}
where the first terms $k=1$ and $k=2$ read
\begin{eqnarray}
\frac{ \partial \langle x \rangle_t   }{\partial t} &&=  v   -  \lambda (1-\gamma) \langle x \rangle_t 
\nonumber \\
\frac{ \partial \langle x^2 \rangle_t   }{\partial t} && = 2 v   \langle x \rangle_t 
-  \lambda (1-\gamma^2) \langle x^2 \rangle_t 
\label{dynmoments12}
\end{eqnarray}
As a consequence, the moments of the steady state $\rho_*(x)$ can be computed recursively
\begin{eqnarray}
\langle x \rangle_* &&=  \frac{ v   } {  \lambda (1-\gamma) }
\nonumber \\
\langle x^2 \rangle_* && = \frac{ 2 v   }  { \lambda (1-\gamma^2) }  \langle x \rangle_*
=  \frac{ 2 v^2   }  { \lambda^2  (1-\gamma) (1-\gamma^2) }
\nonumber \\
...
\nonumber \\
\langle x^k \rangle_* && = \frac{ k v    }
{ \lambda (1-\gamma^k)  }\langle x^{k-1} \rangle_* =  \left( \frac{v}{\lambda} \right)^k \frac{ k!} {\displaystyle \prod_{k'=1}^k (1-\gamma^{k'})}
\label{stmoments}
\end{eqnarray}
and lead to the serie representation of the Laplace transform
\begin{eqnarray}
\hat \rho_*(s) \equiv \int_0^{+\infty} dx  e^{-s x} \rho_*(x) = \sum_{k=0}^{+\infty} \frac{(-s)^k}{k!} \langle x^k \rangle_* 
= \sum_{k=0}^{+\infty}   \frac{ \left( -s \frac{v}{\lambda} \right)^k} {\displaystyle \prod_{k'=1}^k (1-\gamma^{k'})}
\label{laplacemoments}
\end{eqnarray}

The conditions for the existence of the steady state
are analyzed in \cite{blowup} for
the more general models where both the drift $v(x)$ and the jump rate $\lambda(x)$
are polynomial functions of $x$.


\subsection{ Large deviations at Level 2.5  }

Since there is no diffusion $D(x)=0$ and since 
the jump probability is deterministic $ \Pi( x \vert y)  =  \delta \left(x - \gamma y \right)    $,
the large deviations at Level 2.5 simplify (Eqs \ref{ld2.5nodiff} and \ref{ld2.5diffpideter}) 
into
the joint distribution of the empirical density $\rho(.)$
and of the empirical out-flow $Q^-(.)$ with the corresponding density $n$ 
\begin{eqnarray}
 P_T[ \rho(.),n,Q^-(.)] &&  \opsimeq_{T \to +\infty} 
  \delta \left(\int_0^{+\infty} d x \rho(x) -1  \right) \delta \left(  \int_0^{+\infty} d y \ Q^-(y) - n     \right)   
\left[ \prod_{x \geq 0}  \delta \left( v \rho'(x)  + Q^-(x)   -   \frac{1}{\gamma} Q^- \left( \frac{x}{\gamma} \right)  \right) \right]
 \nonumber \\
&& e^{- \displaystyle T   \int_0^{+\infty}  d y
\left[ Q^-(y) \ln \left( \frac{  Q^-(y)  }{    \lambda  \rho(y) }  \right) 
 -  Q^-(y) +   \lambda  \rho(y)  \right] }
\label{ld2.5deterqm}
\end{eqnarray}
while the jump-flow $Q(.,.)$ and the in-flow $Q^+(.)$ 
can be computed from the out-flow $Q^-(.)$
\begin{eqnarray}
Q(x,y) && =  \delta(x- \gamma y) Q^-(y)
\nonumber \\
Q^+(x) && =  \frac{1}{\gamma} Q^- \left( \frac{x}{\gamma} \right)
  \label{Q2etQp}
\end{eqnarray}

One can use the stationarity constraint to eliminate the empirical density $\rho(.)$
in terms of the out-flow $Q^-(.)$
\begin{eqnarray}
  \rho(x)  = \frac{1}{v} \int_x^{\frac{x}{\gamma}} dy Q^-( y) 
  \label{rhoqdeter}
\end{eqnarray}
The normalization of the empirical density becomes
\begin{eqnarray}
\int_0^{+\infty} d x  \rho(x)  = \frac{1}{v} \int_0^{+\infty} dy Q^-( y) \int_{\gamma y}^y dx
= \frac{1-\gamma }{v} \int _0^{+\infty}dy y Q^-( y) 
  \label{normarhoqdeter}
\end{eqnarray}
So the large deviations for the out-flow $Q^-(.)$ and the density $n$ reduce to 
\begin{eqnarray}
 P_T[ n,Q^-(.)] &&  \opsimeq_{T \to +\infty} 
  \delta \left( \frac{1-\gamma }{v} \int _0^{+\infty}dy y Q^-( y) -1  \right) \delta \left(  \int d y \ Q^-(y) - n     \right)   
 \nonumber \\
&& e^{- \displaystyle T  
\left[  -  n +   \lambda   
+ \int_0^{+\infty}  d x 
 Q^-(x) \ln \left( \frac{  Q^-(x)  }{     \frac{\lambda}{v} \int_x^{\frac{x}{\gamma}} dy Q^-( y)  }  \right) 
 \right] }
\label{ld2.5deterqmoutn}
\end{eqnarray}


\subsection{ Large deviations for excursions between jumps }

Here the excursion kernel reduces to Eq. \ref{Wescsansdiff} 
\begin{eqnarray}
W (\tau ,  y  \vert x  )
 =  W_+^{sp}(  y  \vert x  )
  \delta\left( \tau - \int_x^{y}   \frac{ dz }{ v } \right) 
  = W_+^{sp}(  y  \vert x  )
  \delta\left( \tau -   \frac{ y-x  }{ v } \right) 
\label{Wescsansdiffblow}
\end{eqnarray}
with Eq. \ref{WWspatialp} for $y \geq x$
\begin{eqnarray}
W^{sp}_+(  y  \vert x  )
 =   \frac{  \lambda }{ v } \ \ e^{ - \displaystyle \int_x^{y} dz  \frac{ \lambda }{v } }
 =   \frac{  \lambda }{ v } \ \ e^{ - \displaystyle   \frac{ \lambda }{v }  (y-x) }
  \label{WWspatialpblow}
\end{eqnarray}
The empirical excursions contain the same delta function
 for the duration $\tau$ as in Eq \ref{Wescsansdiffblow}
\begin{eqnarray}
q (\tau ,  y  , x  )  = q^{sp}(  y  , x  )  \delta\left( \tau -  \frac{ y-x  }{ v } \right) 
\label{qexcblow}
\end{eqnarray}
where the spatial part $q^{sp}(  y  , x  )$ defined for $y \geq x$
can fluctuate according to the conditional probability of Eq. \ref{ld2.75condi}
\begin{eqnarray}
&& {\cal P}_T^{conditional} [ q^{sp} ( . , .) \vert \rho(.),Q^{-}(.)]
  \opsimeq_{T \to +\infty}  e^{ \displaystyle -T  {\cal I}^{conditional} [ \rho(.), Q^{-}(.),q^{sp}(.,.)] }
\nonumber \\
&& \left[ \prod_{ x>0 }    \delta \left(  \int_x^{+\infty} d y \ q^{sp}(y , x)   - \frac{1}{\gamma} Q^- \left( \frac{x}{\gamma} \right)    \right)  \right]
\left[ \prod_{ y>0 }   \delta \left(   \int_0^y d x \ q^{sp}( y , x)    - Q^-(y)  \right)  \right]
\label{ld2.75condiblow}
\end{eqnarray}
where the conditional rate function (Eq. \ref{rate2.75conditionalhat})
\begin{eqnarray}
  {\cal I}^{conditional} [ \rho(.), Q^{-}(.),q^{sp}(.,.)] 
 = \int_0^{+\infty} d x \int_x^{+\infty} d y \ 
 q^{sp} (y , x)  \ln \left( \frac{ q^{sp} (y , x) }{ 
 \frac{  Q^-( y) }{ \rho(y)  v   } \ \ e^{ -  \int_x^{y} dz  \frac{ Q^-(z) }{ \rho(z) v } }
  \frac{1}{\gamma} Q^- \left( \frac{x}{\gamma} \right)  }   \right) 
  \label{rate2.75conditionalhatblow}
\end{eqnarray}
governs the fluctuations around the optimal value
\begin{eqnarray}
q^{sp}_{opt} (y , x)= \frac{  Q^-( y) }{ \rho(y)  v   } \ \ e^{ -  \int_x^{y} dz  \frac{ Q^-(z) }{ \rho(z) v } }
  \frac{1}{\gamma} Q^- \left( \frac{x}{\gamma} \right)
\label{qspoptimalblow}
\end{eqnarray}
once $[ \rho(.), Q^{-}(.)] $ are given.


\section{ Example of jump-drift process $[ v(x)=-x, \lambda(x)=\lambda ]$with $\Pi(x \vert y)=\theta(x \geq y) \alpha e^{- \alpha (x-y) }$ }

\label{sec_rain}

In this section, we consider the case of the positive jump-drift process with the linear negative drift $v(x)=-x$,
with the uniform jump rate $ \lambda(x)=\lambda $, and where the jump probability describes
positive jumps whose amplitude $z=x-y \geq 0 $ is exponentially distributed
\begin{eqnarray}
 \Pi^{exp}_{\alpha}(x \vert y)= \alpha e^{- \alpha (x-y) } \ \ {\rm for } \ \ x \in [y,+\infty[
\label{rainexp}
\end{eqnarray}
This exponential distribution is often considered in soil moisture models in order to represent rainfall events \cite{waterbalance,daly_rain,daly_rainbis,rainfall}.

 
 \subsection{ Steady state $\rho_*(x)$}

The steady-state $\rho_*(x)$ satisfying Eq. \ref{jumpdiffst}
\begin{eqnarray}
  \frac{ d }{ d x}   \left[ - x \rho_*(x )   \right]
+  \lambda   \rho_*(x) =  \int_0^x d y \alpha e^{- \alpha (x-y) }  \lambda  \rho_*(y)
\label{jumpdiffstzrain}
\end{eqnarray}
is the Gamma-law of shape parameter $\lambda$ and of scale parameter $\frac{1}{\alpha}$
\begin{eqnarray}
 \rho_*(x ) && = \frac{\alpha^{\lambda} }{\Gamma(\lambda) } \ x^{\lambda-1}  e^{\displaystyle - \alpha x }
\label{rainsoluloigamma}
\end{eqnarray}


\subsection{ Large deviations at Level 2.5  }

Since there is no diffusion $D(x)=0$ (Eq. \ref{ld2.5nodiff}) and since the jumps have a positive amplitude $z=x - y \geq 0$,
 the large deviations at Level 2.5 read
\begin{eqnarray}
 P^{2.5[D(.)=0]}_T[ \rho(.), n,Q^{\pm}(.),Q(.,.)]   \opsimeq_{T \to +\infty}  
C^{[D(.)=0]}_{2.5}[ \rho(.), n,Q^{\pm}(.),Q(.,.)] 
e^{- \displaystyle T \left[ I^{[\lambda]}_{2.5} [ \rho(.), Q^-(.)]  
  + I^{[\Pi]}_{2.5} [  Q^-(.),Q(.,.)]  \right]}
\label{ld2.5nodiffrain}
\end{eqnarray}
with the constraints
\begin{eqnarray}
&& C^{[D(.)=0]}_{2.5}[ \rho(.), n,Q^{\pm}(.),Q(.,.)]   =   
 \delta \left(\int_0^{+\infty} d x \rho(x) -1  \right)
\delta \left(  \int_0^{+\infty} d y \ Q^-(y) - n     \right)    \delta \left(  \int_0^{+\infty} d x \ Q^+(x) - n     \right)       
 \label{c2.5nodiffrain}      \\ && 
 \left[  \prod_{x \geq 0  } \delta \left(  \frac{d}{dx} \left[  - x \rho(x) \right] + Q^-(x)   -  Q^+(x)  \right) \right]   
 \left[ \prod_{ x \geq 0 } \delta \left(  \int_0^{x} d y \  Q(x,y) - Q^+(x)     \right)   \right]
\left[ \prod_{ y \geq 0 } \delta \left(  \int_y^{+\infty} d x \ Q(x,y) - Q^-(y) \right) \right]  
 \nonumber   
 \end{eqnarray}
and with the two contributions to the rate function
\begin{eqnarray}
 I^{[\lambda]}_{2.5} [ \rho(.), Q^-(.)]  && = \int_0^{+\infty} d y \left[ Q^-(y) \ln \left( \frac{  Q^-(y)  }{    \lambda   \rho(y) }  \right) 
 -  Q^-(y) +   \lambda   \rho(y)  \right]
\nonumber \\
 I^{[\Pi(x \vert y)  = \alpha e^{- \alpha (x-y) }]}_{2.5} [  Q^-(.),Q(.,.)]  &&= 
 \int_0^{+\infty} d y \int_{x}^{+\infty} d x Q(x,y) \ln \left( \frac{ Q(x,y)  }{   \alpha e^{- \alpha (x-y) } Q^-(y) }  \right) 
\label{rate2.5jumppirain}
\end{eqnarray}

Since the jump probability is given by the exponential form of Eq. \ref{rainexp},
the contraction over the jump flow $Q(.,.)$ can be explicitly computed :
the joint distribution
of $[ \rho(.),n,Q^{\pm}(.)]  $ reads (Eq. \ref{ld2.5diffpicontractionfinalexp})
\begin{eqnarray}
&& P_T^{[\Pi^{exp}_{\alpha}(x \vert y)= \alpha e^{- \alpha (x-y) }]}[ \rho(.), n,Q^{\pm}(.)] 
    \opsimeq_{T \to +\infty}  
 \delta \left(\int_0^{+\infty}  d x \rho(x) -1  \right)
\left[ \prod_{x >0 }  \delta \left( \frac{d}{dx} \left[  - x \rho(x) \right] + Q^-(x)   -  Q^+(x)  \right) \right]     
\nonumber     \\ && 
 \delta \left(  \int_0^{+\infty} d y \ Q^-(y) - n     \right)    \delta \left(  \int_0^{+\infty} d x \ Q^+(x) - n     \right)     
 e^{- \displaystyle T \left( 
I^{[\lambda]}_{2.5} [ \rho(.), Q^-(.)] 
+ I^{exp}_{\alpha} [ Q^+(.), \rho(.) ] 
\right) }
\label{ld2.5diffpicontractionfinalexprain}  
\end{eqnarray}
with the rate function contribution of Eq. \ref{rateexpo} after taking into account $j(x)=  - x \rho(x)$
 \begin{eqnarray}
I^{exp}_{\alpha} [ Q^+(.), \rho(.) ] \equiv    \int_0^{+\infty} dx  \left[  Q^{+}(x) \ln \left( \frac{  Q^+(x)    }{ \alpha  x \rho(x)  }  \right)  
    - Q_+(x) + \alpha  x \rho(x)
    \right]
 \label{rateexporain}
\end{eqnarray}


\subsection{ Large deviations for excursions between jumps }

Here the excursion kernel reduces to Eq. \ref{Wescsansdiff} 
\begin{eqnarray}
W (\tau ,  y  \vert x  )
 =  W^{sp}_-(  y  \vert x  )  \delta\left( \tau - \int_y^{x}   \frac{ dz }{z } \right) 
 = W^{sp}_-(  y  \vert x  )  \delta\left( \tau - \ln  \left( \frac{ x }{y } \right)  \right) 
\label{Wescsansdiffneg}
\end{eqnarray}
with Eq. \ref{WWspatialm} for $0 \leq y \leq x$
\begin{eqnarray}
 W^{sp}_-(  y  \vert x  )
 =    \frac{  \lambda }{ y  } \ \ e^{  \displaystyle - \int_y^x dz  \frac{ \lambda }{ z } } 
 = \frac{  \lambda y^{\lambda-1} }{ x^{\lambda}  }
 \ \ {\rm for } \ \ y \in [0,x]
\label{WWspatialmneg}
\end{eqnarray}

The conditional probability to see spatial excursions
$q^{sp} ( . , .) $ once all the other empirical observables are given reads (Eq. \ref{ld2.75condi})
\begin{eqnarray}
&& {\cal P}_T^{conditional} [ q^{sp} ( . , .) \vert \rho(.), n,Q^{\pm}(.) ] 
\opsimeq_{T \to +\infty}  e^{ \displaystyle -T   {\cal I}^{conditional} [ \rho(.), Q^{\pm}(.),q^{sp}(.,.)]}
  \nonumber \\
&&
 \left[ \prod_{ x \geq 0 }    \delta \left(  \int_0^x d y \ q^{sp}(y , x)   - Q^+(x)     \right)  \right]
\left[ \prod_{ y \geq 0 }   \delta \left(   \int_y^{+\infty} d x \ q^{sp}( y , x)    - Q^-(y)  \right)  \right]
\label{ld2.75condirain}
\end{eqnarray}
where the conditional rate function (Eq. \ref{rate2.75conditionalhat})
\begin{eqnarray}
  {\cal I}^{conditional} [ \rho(.), Q^{\pm}(.),q^{sp}(.,.)] 
 = 
   \int_0^{+\infty} d x \int_0^x d y 
 q^{sp} (y , x)  \ln \left( \frac{ q^{sp} (y , x) }{ 
 \frac{  Q^-( y) }{ y \rho(y)   } \ \ e^{ -  \int_y^{x} dz  \frac{ Q^-(z) }{ z \rho(z)  } }  Q^+(x)  }   \right)
  \label{rate2.75conditionalhatrain}
\end{eqnarray}
governs the fluctuations around the optimal value
\begin{eqnarray}
q^{sp}_{opt} (y , x)= \frac{  Q^-( y) }{ y \rho(y)   } \ \ e^{ -  \int_y^{x} dz  \frac{ Q^-(z) }{ z \rho(z)  } }  Q^+(x)
\label{qspoptimalrain}
\end{eqnarray}
once $[ \rho(.), Q^{\pm}(.)] $ are given.


\section{ Example : jump-diffusion $[D(x)=D, v(x)=0, \lambda(x)=\lambda ]$ with resetting $\Pi^{reset}( x \vert y)  =  R(x) $ }

\label{sec_diffusionresetR}

As an example of stochastic resetting (see the review \cite{review_reset} and references therein)
towards an arbitrary probability distribution $R(x)$ (Eq. \ref{reset})
\begin{eqnarray}
\Pi^{reset}( x \vert y)  =  R(x) 
\label{resetex}
\end{eqnarray}
instead of the resetting towards the origin discussed in section \ref{sec_reset0},
let us consider the jump-diffusion process without drift $v(x)=0$,
with uniform diffusion coefficient $D(x)=D$ and uniform jump rate $ \lambda(x)=\lambda $.

 
 \subsection{ Steady state $\rho_*(x)$}

The steady-state $\rho_*(x)$ satisfying Eq. \ref{jumpdiffst}
\begin{eqnarray}
  -D    \frac{ d^2  \rho_*(x)}{ d^2 x}   +  \lambda   \rho_*(x) = \lambda R(x) 
\label{jumpdiffstreset}
\end{eqnarray}
can be written as
\begin{eqnarray}
 \rho_*(x ) =   \int_{-\infty}^{+\infty} dx_0 G(x,x_0) R(x_0) 
\label{pgreen}
\end{eqnarray}
where the Green function $G(x,x_0)$ satisfying
\begin{eqnarray}
-D    \frac{ d^2  G(x,x_0)}{ d x^2}   
+  \lambda   G(x,x_0) =\lambda \delta(x-x_0)
\label{greeneq}
\end{eqnarray}
corresponds to the elementary solution associated to the deterministic resetting towards $x_0$.
The solution that is well-behaved at $x \to \pm \infty$ 
\begin{eqnarray}
 G(x,x_0) =    \frac{1}{ 2  } \sqrt{ \frac{\lambda}{D} } e^{ \displaystyle - \vert x - x_0 \vert \sqrt{ \frac{\lambda}{D} }}
 \label{greeneqsimplesol}
\end{eqnarray}
yields the steady state (Eq. \ref{pgreen})
\begin{eqnarray}
 \rho_*(x )  
  = \frac{1}{ 2  } \sqrt{ \frac{\lambda}{D} }
\left[ e^{ \displaystyle -x  \sqrt{ \frac{\lambda}{D} }}
 \int_{-\infty}^x dx_0  R(x_0)  e^{ \displaystyle x_0   \sqrt{ \frac{\lambda}{D} }}
 +  e^{ \displaystyle x  \sqrt{ \frac{\lambda}{D} }}
 \int_x^{+\infty} dx_0  R(x_0)  e^{ \displaystyle -x_0 \sqrt{ \frac{\lambda}{D} }} \right]
\label{pgreensimple}
\end{eqnarray}


\subsection{ Large deviations at Level 2.5  }

For the present model, the large deviations at Level 2.5 of Eq. \ref{ld2.5diff}
read with the constraints $C_{2.5}[ \rho(.), j(.),Q(.,.)] $ given in Eq. \ref{c2.5}
\begin{eqnarray}
&& P^{2.5}_T[ \rho(.), j(.),n,Q^{\pm}(.),Q(.,.)]   \opsimeq_{T \to +\infty}  
C_{2.5}[ \rho(.), j(.),n,Q^{\pm}(.),Q(.,.)] 
\nonumber       \\ && 
e^{- \displaystyle T \left[ I_{2.5}^{[D]} [ \rho(.), j(.)]  
  +  I^{[\lambda]}_{2.5} [ \rho(.), Q^-(.)]  
  + I^{[\Pi]}_{2.5} [  Q^-(.),Q(.,.)] \right]}
\label{ld2.5diffresetR}
\end{eqnarray}
with the three contributions to the rate function
\begin{eqnarray}
  I_{2.5}^{[D]} [ \rho(.), j(.)]  
&& = \int_{-\infty}^{+\infty} \frac{d x}{ 4 D \rho(x) } \left[ j(x) +D \rho' (x) \right]^2
 \nonumber \\
 I^{[\lambda]}_{2.5} [ \rho(.), Q^-(.)]  && 
 = \int_{-\infty}^{+\infty} d y \left[ Q^-(y) \ln \left( \frac{  Q^-(y)  }{    \lambda   \rho(y) }  \right) 
 -  Q^-(y) +   \lambda   \rho(y)  \right]
 \nonumber \\
 I^{[\Pi]}_{2.5} [  Q^-(.),Q(.,.)] && =  \int d x \int d y Q(x,y) \ln \left( \frac{ Q(x,y)  }{  R(x) Q^-(y) }  \right) 
\label{rate2.5troisresetR}
\end{eqnarray}

Since the jump probability corresponds to the stochastic resetting form of Eq. \ref{resetex},
the contraction over the jump flow $Q(.,.)$ can be explicitly computed :
the joint distribution
of $[ \rho(.),n,Q^{\pm}(.)]  $ reads (Eq \ref{ld2.5diffpicontractionfinal}) reads
\begin{eqnarray}
&& P_T^{\Pi^{reset}( x \vert y) =R(x)}[ \rho(.), j(.),n,Q^{\pm}(.)] 
    \opsimeq_{T \to +\infty}  
 \delta \left(\int d x \rho(x) -1  \right)
\left[ \prod_{x }  \delta \left(  j'(x) + Q^-(x)   -  Q^+(x)  \right) \right]     
\label{ld2.5diffpicontractionfinalreset}      \\ && 
 \delta \left(  \int d y \ Q^-(y) - n     \right)    \delta \left(  \int d x \ Q^+(x) - n     \right)     
 e^{- \displaystyle T \left( 
I_{2.5}^{[D]} [ \rho(.), j(.)]  
+I^{[\lambda]}_{2.5} [ \rho(.), Q^-(.)] 
+  I^{[R]} [ n, Q^+(.)] 
\right) }
\nonumber 
\end{eqnarray}
with the last contribution of the rate function
 \begin{eqnarray}
I^{[R]} [ n, Q^+(.)] && \equiv    \int_{-\infty}^{+\infty} d x  Q^+(x) \ln \left( \frac{Q^+(x)  }{ n R(x) } \right) 
 \label{ratereset3}
\end{eqnarray}


\subsection{ Large deviations for the jumps and for the excursions between jumps }

Since the jump rate is uniform $\lambda(x)=\lambda$, 
the excursion kernel of Eq. \ref{Wescfactor} is factorized
\begin{eqnarray}
W(\tau ,  y  \vert x  )
 = E^{exc} (\tau)  P^{free}_{\tau}(  y  \vert x  )
\label{Wescfactorresetdiff}
\end{eqnarray}
into the normalized exponential probability to see the duration $\tau \in ]0,+\infty[$
\begin{eqnarray}
E^{exc} (\tau) = \lambda  e^{\displaystyle - \tau \lambda}  
\label{tauexpobis}
\end{eqnarray}
and into the gaussian free propagator (uniform diffusion $D(y)=D$ without drift $v(y)=0$)
\begin{eqnarray}
P^{free}_{\tau}(  y  \vert x  )
 =    \frac{1}{ \sqrt{ 4 \pi D \tau } }e^{ \displaystyle - \frac{ \left(y - x  \right)^2 }{4D \tau } }
\label{gauss}
\end{eqnarray}

The large deviations for the empirical jumps and for the empirical excursions between jumps of Eq. \ref{ld2.75}
read with the constraints $ {\cal C}[ n; Q^{\pm}(.); Q(.,.) ; q(.) ; q(.,.,.)  ] $ given in Eq. \ref{c2.75}
\begin{eqnarray}
 {\cal P}_T[ n; Q^{\pm}(.) ;  Q(.,.) ; q(.) ; q(.,.,.)   ] && \opsimeq_{T \to +\infty}  
{\cal C}[ n; Q^{\pm}(.); Q(.,.) ; q(.) ; q(.,.,.)  ] 
\nonumber \\
&& e^{\displaystyle - T \left[  \int d x \int d y
Q(x,y) \ln \left( \frac{ Q(x,y)  }{  R(x)  Q^-(y) }  \right)  +  {\cal I}^{[W]} [  Q^{+}(.) ; q(.,.,.)  ] \right]   }
\label{ld2.75reset}
\end{eqnarray}
where the rate function contribution involving the excursion kernel $W(\tau,  y  \vert x  )$ of Eq. \ref{Wescfactorresetdiff} reads
\begin{eqnarray}
 {\cal I}^{[W]} [  Q^{+}(.) ; q(.,.,.)  ] 
= \int_0^{+\infty} d \tau \int d x \int d y \ 
 q(\tau,y , x)  \ln \left( \frac{ q(\tau,y , x) }{    
   \frac{\lambda}{ \sqrt{ 4 \pi D \tau } }e^{ - \tau \lambda - \frac{ \left(y - x  \right)^2 }{4D \tau } }
  Q^+(x)  }   \right)
\label{rate2.75excreset}
\end{eqnarray}


\section{ Example of jump-diffusion $[D(x)=D, v(x)=-x, \lambda(x)=\lambda ]$ with $\Pi( x \vert y)  = H(x-y)  $  }

\label{sec_OUavecjumpHz}

As last example, let us consider the case of uniform diffusion $D(x)=D $
with the linear drift towards the origin $v(x)=-x $,
with uniform jump rate $ \lambda(x)=\lambda $,
while the jump probability involves an arbitrary function $H(z)$ of the amplitude $z=x-y$
\begin{eqnarray}
\Pi( x \vert y)  =  H(x-y)
\label{difference}
\end{eqnarray}

 
 \subsection{ Steady state $\rho_*(x)$ and its Fourier transform ${\hat \rho}_*(k)  $ }

Eq. \ref{jumpdiffst} for the steady-state $\rho_*(x)$
\begin{eqnarray}
 -D    \frac{ d^2  \rho_*(x)}{ d^2 x} -  \frac{ d }{ d x}   \left[ x \rho_*(x )     \right]
+  \lambda   \rho_*(x) = \lambda  \int d y  H( x - y)    \rho_*(y)
\label{jumpdiffstrestoring}
\end{eqnarray}
can be translated in terms of the Fourier transforms
\begin{eqnarray}
{\hat \rho}_*(k) && \equiv \int_{-\infty}^{+\infty} dx e^{ikx}  \rho_*(x ) 
\nonumber \\
{\hat H}(k) && \equiv \int_{-\infty}^{+\infty} dz e^{ikz}  H(z ) 
\label{fourier}
\end{eqnarray}
into the first-order differential equation in $k$
\begin{eqnarray}
   \frac{ d {\hat \rho}_*(k)}{ d k}  
 = - D k {\hat \rho}_*(k) + \lambda  \left( \frac{ {\hat H }(k) -1}{k}  \right)  {\hat \rho}_*(k)
\label{eqdifffourier}
\end{eqnarray}
Using the normalization of the steady state
\begin{eqnarray}
{\hat \rho}_*(k=0) = \int_{-\infty}^{+\infty} dx   \rho_*(x ) =1
\label{fourierz}
\end{eqnarray}
the solution of Eq. \ref{eqdifffourier} reads
\begin{eqnarray}
   {\hat \rho}_*(k)   
 = e^{\displaystyle  - \frac{D k^2}{2}  -  \lambda  \int_0^k dk' \left( \frac{ 1- {\hat H }(k') }{k'}  \right)  } 
\label{fouriersolu}
\end{eqnarray}


\subsubsection { Example of symmetric L\'evy jumps $H(z)=L_{\mu}(z)$ of index $\mu \in ]0,2[$}

An interesting example is when the size $z=x-y$ of the jump is drawn with the L\'evy symmetric stable law $H(z)=L_{\mu}(z)$
of index $\mu \in ]0,2[$ and of characteristic scale $\Delta$
\begin{eqnarray}
H(z) =L_{\mu}(z) = \int_{-\infty}^{+\infty} \frac{dk}{2 \pi} e^{ - i k z - \Delta^{\mu} \vert k \vert^{\mu} } 
\label{levy}
\end{eqnarray}
displaying the power-law decay of exponent $(1+\mu)$
\begin{eqnarray}
L_{\mu}(z) \opsimeq_{ z \to \pm \infty}  \frac{ \Gamma(1+ \mu) \sin \left( \frac{\pi \mu}{2}\right) \Delta^{\mu}}{ \pi \vert z \vert^{1+\mu}}
\label{levypower}
\end{eqnarray}
For instance, the value $\mu=1$ corresponds to the Cauchy distribution 
\begin{eqnarray}
L_{\mu=1}(z) = \int_{-\infty}^{+\infty} \frac{dk}{2 \pi} e^{ - i k z - \Delta \vert k \vert } = \frac{\Delta}{\pi ( z^2+\Delta^2) }
\label{cauchy}
\end{eqnarray}
Then the steady state of Eq. \ref{fouriersolu} 
\begin{eqnarray}
   {\hat \rho}_*(k)   
 = e^{\displaystyle  - \frac{D k^2}{2}  
 -  \lambda  \int_0^k dk' \left( \frac{ 1- e^{  - \Delta^{\mu} \vert k' \vert^{\mu}}}{k'}  \right)  } 
 = e^{\displaystyle  - \frac{D k^2}{2}  
 -  \lambda  \int_0^{\vert k \vert}  dk' \left( \frac{ 1- e^{  - \Delta^{\mu} (k')^{\mu}}}{k'}  \right)  } 
\label{jumpdiffstrestoringfouriersolulevy}
\end{eqnarray}
inherits the L\'evy singularity in $\vert k \vert^{\mu} $ near the origin $k \to 0$
\begin{eqnarray}
   {\hat \rho}_*(k)   \opsimeq_{k \to 0} 1 
 - \frac{ \lambda \Delta^{\mu} }{\mu} \vert k \vert^{\mu}  
\label{singlevy}
\end{eqnarray}
so that the steady state $\rho_*(x)$ decays only as the power-law $ \vert x \vert^{-1-\mu}$ 
in real space $x \to \pm \infty$.


\subsubsection { Example of symmetric exponential jumps $H(z)=\frac{\alpha}{2} e^{- \alpha \vert z \vert }$ }

When the size $z=x-y$ of the jump is drawn with the symmetric exponential distribution
\begin{eqnarray}
H(z) = \frac{\alpha}{2} e^{- \alpha \vert z \vert }
\label{expsym}
\end{eqnarray}
its Fourier transform
\begin{eqnarray}
{\hat H}(k) = \frac{1}{ 1+ \frac{k^2}{ \alpha^2} } 
\label{expsymfourier}
\end{eqnarray}
yields that Eq. \ref{fouriersolu} becomes
\begin{eqnarray}
   {\hat \rho}_*(k)   
 = e^{\displaystyle  - \frac{D k^2}{2}  -  \frac{\lambda}{2}  \ln \left( 1+ \frac{k^2}{ \alpha^2}\right)  } 
 = \frac{ e^{\displaystyle  - \frac{D k^2}{2} } }{\left( 1+ \frac{k^2}{ \alpha^2}\right)^{ \frac{\lambda}{2}} } 
\label{symexpsolu}
\end{eqnarray}


\subsection{ Large deviations at Level 2.5  }

For the present model, the large deviations at Level 2.5 of Eq. \ref{ld2.5diff}
read with the constraints $C_{2.5}[ \rho(.), j(.),Q(.,.)] $ given in Eq. \ref{c2.5}
\begin{eqnarray}
 P^{2.5}_T[ \rho(.), j(.),n,Q^{\pm}(.),Q(.,.)]   \opsimeq_{T \to +\infty}  
C_{2.5}[ \rho(.), j(.),Q(.,.)] 
e^{- \displaystyle T \left[ I_{2.5}^{[D]} [ \rho(.), j(.)]  
  +  I^{[\lambda]}_{2.5} [ \rho(.), Q^-(.)]  
  + I^{[\Pi]}_{2.5} [  Q^-(.),Q(.,.)] \right]}
\label{ld2.5diffOU}
\end{eqnarray}
with the three contributions to the rate function
\begin{eqnarray}
 I_{2.5}^{[D,v]} [ \rho(.), j(.)]  
&& = \int \frac{d x}{ 4 D \rho(x) } \left[ j(x) +x  \rho(x) +D \rho' (x) \right]^2
\nonumber \\
 I^{[\lambda]}_{2.5} [ \rho(.), Q^-(.)] 
 &&  = \int d y \left[ Q^-(y) \ln \left( \frac{  Q^-(y)  }{    \lambda   \rho(y) }  \right) 
 -  Q^-(y) +   \lambda   \rho(y)  \right]
\nonumber \\
 I^{[\Pi]}_{2.5} [  Q^-(.),Q(.,.)] && = 
 \int d x \int d y Q(x,y) \ln \left( \frac{ Q(x,y)  }{  H(x-y) Q^-(y) }  \right) 
\label{rate2.5jumppiOU}
\end{eqnarray}


\subsection{ Large deviations for the empirical jumps and for the empirical excursions between jumps }

Since the jump rate is uniform $\lambda(x)=\lambda$, 
the excursion kernel of Eq. \ref{Wescfactor} is factorized
\begin{eqnarray}
W(\tau ,  y  \vert x  )
 = E^{exc} (\tau)  P^{free}_{\tau}(  y  \vert x  )
\label{WescfactorOU}
\end{eqnarray}
into the normalized exponential probability to see the duration $\tau \in ]0,+\infty[$
\begin{eqnarray}
E^{exc} (\tau) = \lambda  e^{\displaystyle - \tau \lambda}  
\label{tauexpoter}
\end{eqnarray}
and into the free Ornstein-Uhlenbeck propagator ($[D(x)=D, v(x)=-x]$)
\begin{eqnarray}
 P^{free}_{\tau}(  y  \vert x  ) = 
  \frac{1}{ \sqrt{ 2 \pi D (1- e^{-2\tau}) } }e^{ \displaystyle - \frac{ \left(y - x e^{-\tau}  \right)^2 }{2 D (1- e^{-2\tau})  } }
\label{OUpropagateur}
\end{eqnarray}

The large deviations for the empirical jumps and for the empirical excursions between jumps of Eq. \ref{ld2.75}
read with the constraints $ {\cal C}[ n; Q^{\pm}(.); Q(.,.) ; q(.) ; q(.,.,.)  ] $ given in Eq. \ref{c2.75}
\begin{eqnarray}
&& {\cal P}_T[ n; Q^{\pm}(.) ;  Q(.,.) ; q(.) ; q(.,.,.)   ] \opsimeq_{T \to +\infty}  
{\cal C}[ n; Q^{\pm}(.); Q(.,.) ; q(.) ; q(.,.,.)  ] 
\nonumber \\
&&  e^{\displaystyle - T \left[  \int d x \int d y
Q(x,y) \ln \left( \frac{ Q(x,y)  }{   H(x-y)  Q^-(y) }  \right)
+{\cal I}^{[W]} [  Q^{+}(.) ; q(.,.,.)  ]
\right]  }
\label{ld2.75OU}
\end{eqnarray}
where the rate function contribution involving the excursion kernel $W(\tau,  y  \vert x  )$ of Eq. \ref{WescfactorOU} reads
\begin{eqnarray}
 {\cal I}^{[W]} [  Q^{+}(.) ; q(.,.,.)  ] 
= \int_0^{+\infty} d \tau \int d x \int d y \ 
 q(\tau,y , x)  \ln \left( \frac{ q(\tau,y , x) }{    
  \frac{\lambda}{ \sqrt{ 2 \pi D (1- e^{-2\tau}) } }e^{ - \tau \lambda - \frac{ \left(y - x e^{-\tau}  \right)^2 }{2 D (1- e^{-2\tau})  } }
  Q^+(x)  }   \right)
\label{rate2.75excOU}
\end{eqnarray}


\section{ Conclusion  }

\label{sec_conclusion}

In this paper, we have considered one-dimensional Jump-Drift and Jump-Diffusion processes,
defined in terms of four space-dependent parameters, namely the drift $v(x)$, the diffusion coefficient $D(x)$, the jump rate $\lambda(x)$ and the jump probability $\Pi(x' \vert x)$.
We have assumed that these parameters produce some normalizable steady state and we have analyzed
the large deviations of a long dynamical trajectory from two points of view.
We have first applied the Large deviations at Level 2.5 to the joint probability of the empirical time-averaged density $\rho(x)$, of the empirical time-averaged current $j(x)$ and of the empirical time-averaged jump-flow $Q(x,y)$.
We have then focused on the alternate Markov chain that governs the series of all the jump events of a long trajectory
in order to obtain the large deviations at Level 2.5 for the joint probability of the empirical jumps and of the empirical excursions between consecutive jumps. Finally, we have applied these two general frameworks to three examples of positive jump-drift processes without diffusion, and to two examples of jump-diffusion processes, in order to illustrate  various simplifications that may occur in rate functions and in contraction procedures.


\end{document}